\documentclass[10pt,preprint]{aastex}

\newcommand{\Ni}{{$^{56}$Ni}}
\newcommand{\Co}{$^{56}$Co}
\newcommand{\Fe}{$^{56}$Fe}
\newcommand{\Ms}{$M_{\odot}$}
\newcommand      \grays       {$\gamma$-rays}
\newcommand{\arp}{Ar$^+$}
\newcommand{\hep}{He$^+$}
\newcommand{\nep}{Ne$^+$}

\newcommand{\env}{$\sim$}
\newcommand{\sif}{Mg$_2$SiO$_4$}
\newcommand{\sie}{MgSiO$_3$}
\newcommand{\sili}{SiO$_2$}
\newcommand{\al}{Al$_2$O$_3$}
\newcommand{\magn}{Fe$_3$O$_4$}
\newcommand{\mic}{$\mu$m}

\shorttitle{DUST SYNTHESIS IN TYPE II-P SUPERNOVAE}
\shortauthors{Sarangi \& Cherchneff }

\begin{document}


\title{THE CHEMICALLY-CONTROLLED SYNTHESIS OF DUST \\ IN TYPE II-P SUPERNOVAE}

\author{Arkaprabha Sarangi \& Isabelle Cherchneff \altaffilmark{1}}

\altaffiltext{1}{Departement Physik, Universit{\"a}t Basel, CH-4056 Basel, Switzerland; arkaprabha.sarangi@unibas.ch; isabelle.cherchneff@unibas.ch}


\begin{abstract}

We study the formation of molecules and dust clusters in the ejecta of solar metallicity, Type II-P supernovae using a chemical kinetic approach. We follow the evolution of molecules and small dust cluster masses from day 100 to day 1500 after explosion. We consider stellar progenitors with initial mass of 12, 15, 19 and 25  \Ms~that explode as supernovae with stratified ejecta. The molecular precursors to dust grains comprise molecular chains, rings and small clusters of silica, silicates, metal oxides, sulphides and carbides, pure metals, and carbon, where the nucleation of silicate clusters is described by a two-step process of metal and oxygen addition. We study the impact of the \Ni~mass on the type and amount of synthesised dust.

We predict that  large masses of molecules including CO, SiO, SiS, O$_2$, and SO form in the ejecta. We show that the discrepancy between the small dust masses detected at infrared wavelengths some 500 days post-explosion and the larger amounts of dust recently detected with Herschel in supernova remnants can be explained by the non-equilibrium chemistry linked to the formation of molecules and dust clusters in the ejected material. Dust gradually builds up from small (\env\ $10^{-5}$ \Ms) to large masses ($\sim 5\times 10^{-2}$ \Ms) over a 5 yr period after explosion. Subsequent dust formation and/or growth is hampered by the shortage of chemical agents participating in the dust nucleation and the long time scale for accretion. The results highlight the dependence of the dust chemical composition and mass on the amount of \Ni~synthesised during the explosion. This dependence may partly explain the diversity of epochs at which dust forms in supernovae. More generally, our results indicate that type II-P supernovae are efficient but moderate dust makers with an upper limit on the mass of synthesised dust ranging from \env\ 0.03 to 0.09 \Ms. Other dust sources must then operate at high redshift to explain the large quantities of dust present in young galaxies in the early universe.

\end{abstract}



\section{Introduction}
Stars with an initial mass on the main sequence comprised between 8 \Ms~and 30 \Ms~usually end their life as Type II-P supernovae (hereafter SNe). Despite the huge amount of energy released by the explosion ($\sim 1\times 10^{51}$ erg), and the harsh physical conditions that characterise the ejected stellar gas (hereafter referred as the ejecta), dust and molecules have been detected in many SNe a few hundred days after the explosive event. The first evidence for dust synthesis in a SN ejecta was brought with the explosion of SN1987A in the Large Magellanic Cloud more than twenty five years ago. The extensive observational coverage of the event at mid-infrared (IR) wavelengths allowed the detection of the fundamental and overtone transitions of a few molecules, specifically carbon monoxide, CO, and silicon monoxide, SiO, as early as $\sim$120 days post-explosion. The observations also highlighted the formation of dust grains after day 400 \citep{spi88,luc89,meik89,mos89,roch91,dan91,wood93}. 

Since then, warm dust has been detected in several SNe \citep[e.g.][]{elma03, ko05,ko06,ko09,sug06,ins11,gal12}. An excess in the mid-IR, combined with a decrease of several magnitudes in the optical light curve, and blue-shifted emission lines are the usual indicators of the synthesis of dust in the ejecta. The fundamental band of SiO has been detected in a few SNe, e.g., SN2004et, and the gradual fading of the transition over time was ascribed to the depletion of SiO in the condensation process of dust grains in the ejecta \env\ 400 days post-outburst \citep{ko09}. Most important are the small amounts of warm dust derived from modelling the mid-IR excess in SNe with masses that range from $1\times 10^{-5}$ \Ms~to $1\times 10^{-3}$ \Ms. These values are usually derived assuming a homogenous ejecta and a mixture of silicates and carbon, the prevalent types of dust in galaxies, while other condensates such as metal sulphides and oxides may be present in SN ejecta \citep{cher10a}. Larger dust masses arise from the assumption of a clumpy ejecta \citep{erco07}, but the final values always remain small between 200 and 600 days post-outburst. These results do not support the hypothesis that SNe are important dust contributors to galaxies locally and at high redshift. If SN explosions were to provide the large amounts of dust needed to reproduce the reddening of distant quasars and metal measurements in damped Ly$\alpha$ systems \citep{pei91, pet94, ber03}, the dust yield per SN needs to be as high as $\sim$ 1 \Ms~\citep{dwek07}. Such a high value is difficult to reconcile with the small masses of warm dust detected in the IR. 

The latest data on SN remnants (SNRs) obtained with the submilimetre (submm) Herschel telescope have cast a new light on the dust released by SN events. A large mass of cold ejecta dust amounting to 0.08 \Ms~ has been derived in the 330 yr old SNR Cas A \citep{bar10, sib10}. In the Crab Nebula, a 1050 yr old pulsar wind SNR, cool dust was recently detected in the filaments and the derived dust masses amount to 0.1 $-$ 0.24 \Ms, depending on the type of dust assumed \citep{gom12}. Finally, $0.4-0.7$ \Ms~of cool, ejecta dust have been inferred from submm flux data in the young remnant SN1987A \citep{mat11}. These cold dust masses are large compared with those derived from IR observations, and imply that either dust grains continue to form in the SN remnant decades after their initial condensation at day $\sim$ 400, or the IR observations only probe the dust content of the ejecta at early time when the ejecta dust may still form at later epochs in the nebular phase. The first scenario is unlikely because high gas temperatures are required to overcome the activation energy barriers characterising the neutral processes involved in the nucleation of dust, and large densities are also necessary to ensure the efficiency of these reactions \citep{cher10b}. These two conditions are not met in the SNR gas. 

In the present paper, we report on new physico-chemical models of the stratified ejecta of SNe with different progenitor masses and solar metallicity. We study the formation of molecules and small molecular clusters implicated in the nucleation phase of the synthesis of dust, and describe the different steps involved in dust nucleation following a chemical kinetic approach, following previous studies of the chemistry of primeval SNe \citep{cher08, cher09,cher10a}. We include the new nucleation chemistry of small silicate clusters as described by \citet{gou12}. These clusters set an upper limit to the final dust mass formed since they represent a bottleneck to the condensation phase of dust. In \S2, we describe the physical and chemical model of stratified ejecta. The results on elements, molecules and dust clusters are presented in \S3, where we discuss the impact of the \Ni\ mass and compare our results to existing studies. A discussion follows in \S4. 

\section{THE PHYSICAL AND CHEMICAL MODEL}

The helium core of a massive star exploding as a supernova is crossed by a blast wave that deposits energy in the gas. When encountering the progenitor envelope, this wave triggers a reverse shock at the base of the envelope that propagates inward and produces Rayleigh-Taylor instabilities and macroscopic mixing in the helium core. The mixing ceases after a few days \citep{jog10} and the partial fragmentation of the helium core proceeds with time. Radioactive \Ni~decays into \Co~on a time scale of a few days. In turn, \Co~decays into \Fe~with a half-life of $\sim$ 113 days, creating a flux of $\gamma$~photons that pervades the ejecta. The degradation of \grays~to X-rays and ultraviolet (UV) photons occurs by Compton scattering and creates a population of fast Compton electrons in the ejecta. These fast electrons ionise the gas, and produce ions such as \arp, \nep, and \hep,~that are key species in the ejecta chemistry.  The physical models of the stratified ejecta are presented in the next section, followed by a section on the chemistry. 

\subsection{Physical Model}

Stratified ejecta are considered for massive stellar progenitors of masses 12, 15, 19, and 25 \Ms. This choice of progenitor masses is based on the availability of SN nucleosynthesis models in the literature, and corresponds to values derived from the estimate of \Ni\ mass from the SN light curve. Most Type II-P SNe originate from the explosion of massive stars with typical masses of 12 \Ms\ $-$ 15 \Ms. In the cases of SN1987A and Cas A, a progenitor of mass $\sim$ 19 \Ms~has been inferred \citep{woos88,krau08}. The most massive progenitor, 25 \Ms, is considered as a surrogate of the massive SNe characterising the explosion of Population II stars at high redshift \citep{tum06}, whereas the 12 \Ms\ progenitor represents low-mass SNe, including some members of the faint SN subclass characterised by a low mass progenitor ($8 -10$ \Ms), and a low mass of processed \Ni. In these faint SNe, e.g., SN2011ht \citep{mau12}, dust forms as early as $\sim$ 100 days after explosion. 

The stratified ejecta is described by the mass zones of the progenitor core given by the explosion models, and we assume that the gas within each zone is fully-microscopically mixed. No gas leakage between different zones is assumed. The initial chemical composition of the ejecta in the form of the elemental mass yields are taken from \citet{rau02} for the 15, 19, and 25 \Ms~progenitors, while that of the 12 \Ms~progenitor is from \citet{woos07}. The elemental mass yields for all progenitors are summarised in table \ref{tbl-1}. 

Synthetic ejecta temperature and number density profiles were constructed based on the explosion model for a Type II-P supernova with a 17 \Ms~progenitor provided by \citet{noz10}. For  the sake of simplicity, we choose this model for all SN progenitor studied, and the gas parameters are listed in table \ref{tbl-2} for the 15 \Ms\ progenitor as a function of post-explosion time and ejecta zoning. The temperature variation as a function of post-explosion time is given by 
\begin{equation}
\label{temp}
T_{gas}(M_r,t) = T_{gas}(M_r,100)\times (t/100)^{-1.26},
\end{equation}
where T$_{gas}(M_r, 100)$ is the gas temperature 100 days after explosion, $M_r$ is the mass coordinate, and $t$ is the time. In the explosion model of \citet{noz10}, the gas temperature varies with the mass coordinate over the ejecta owing to differential deposition of energy in the helium core. We then assume different initial temperatures with mass zones at 100 days for all progenitor models; the initial T$_{gas}(M_r, 100)$ values are given in table \ref{tbl-2}. 

Assuming homologous expansion, the gas density is given by 
\begin{equation}
\label{dens}
n_{gas}(M_r,t) = \rho_{gas}(M_r,100)/\mu(M_r,t)\times (t/100)^{-3}. 
\end{equation}
where $\rho_{gas}(M_r,100)$ and $\mu(M_r, t)$  are the gas density 100 days post-outburst and the gas mean molecular weight at time $t$ in the mass zone of coordinate $M_r$, respectively. According to the gas density profiles in Figure 2 of \citet{noz10}, we assume a constant, initial gas density $\rho_{gas}(100)$ independent of mass coordinate for all progenitor masses, with $\rho_{gas}(100)=1.1 \times 10^{-11}$ g cm$^{-3}$. All progenitor masses are characterised by an explosion energy of $1 \times 10^{51}$ erg, while the effective $\gamma$-ray optical depth at 100 days $\tau(100)$ has been estimated according to Cherchneff \& Dwek (2009) and are 13.5, 17.5, 23, and 29 for the 12, 15, 19, and 20 \Ms\ progenitor, respectively.%

\subsection{Chemical Model}
\label{chem}

The various atoms, molecules, and ions assumed to form in the SN ejecta and considered in our chemical scheme are summarised in table \ref{tbl-3}. We model the chemistry in the ejecta considering all possible types of chemical reactions relevant to hot and dense environments. All chemical pathways that lead to the formation of linear molecules, carbon chains and rings, and small dust clusters include neutral-neutral processes such as termolecular, bimolecular, and radiative association reactions, and charge exchange reactions, whereas destruction is described  by thermal fragmentation, neutral-neutral processes (i.e., oxidation reactions of carbon chains and all reverse processes of the formation reactions), ion-molecule recombination processes and charge exchange reactions. 

\begin{figure*}
\figurenum{1}
\epsscale{1.}
\plotone{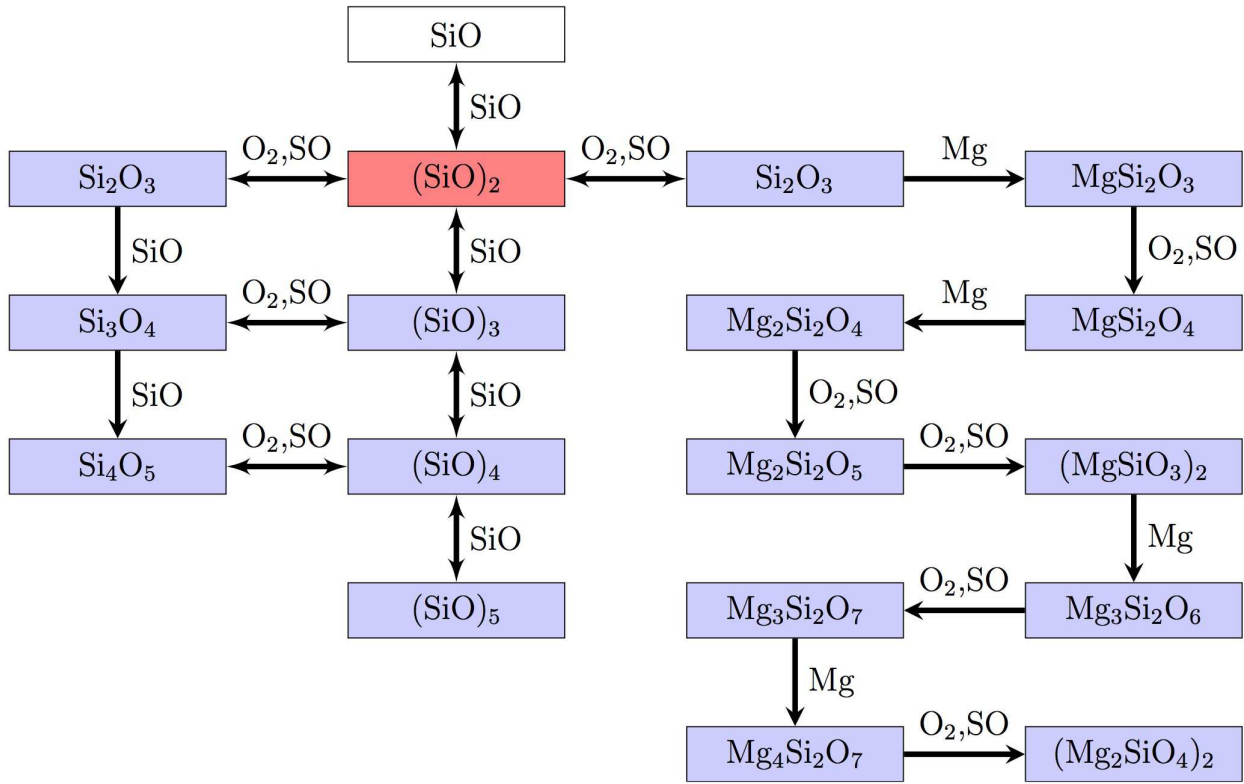}
\caption{Chemical nucleation processes involved in the formation of enstatite and forsterite dimers (Mg$_2$Si$_2$O$_6$ and Mg$_4$Si$_2$O$_8$, respectively) according to \citet{gou12}. Reactant species are given for each process.\label{fig1}}
\end{figure*}

The nucleation scheme to silica and silicate clusters is illustrated in Figure \ref{fig1} and the full chemical network describing the nucleation of silicate clusters is listed in table \ref{tbl-A1} in the Appendix. Small silica clusters form according to the processes described by \citet{cher10a}, where the study of SiO dimerisation by \citet{zac93} was used. In the present study, we consider as "silica", the ensemble of (SiO)$_n$ clusters that form, as these small clusters will condense to form amorphous silica. A possible disproportionation into SiO$_2$ and Si$_2$ components in the condensed amorphous compound is to be expected \citep{reb08}, but we ignore a separation of these two phases in the present study. The description of the growth pathways of small silicate clusters, namely forsterite dimer (Mg$_2$SiO$_4$)$_2$ and enstatite dimer (MgSiO$_3$)$_2$, is based on the work by \citet{gou12}. This study indicates possible chemical routes to the formation of the silicate dimers involving the formation of the SiO dimer (SiO)$_2$ ring and its growth to Si$_2$O$_3$ through the reaction with O$_2$ and SO. The subsequent pathway involves the addition of a Mg atom into the Si$_2$O$_3$ structure. The later growth of clusters is described by one oxygen-addition step followed by one Mg inclusion as a recurrent growth scenario. We consider different oxidising agents, including atomic O, O$_2$ and SO, and find that reactions with O$_2$ and SO are prevalent. Atomic oxygen is very abundant in the O-rich zones 1B and 2, but its inclusion in clusters proceeds through slow reactions such as termolecular (cluster + O + M $\rightarrow$ [cluster+O] + M) and radiative association (cluster + O $\rightarrow$ [cluster+O] + h$\nu$) processes. Both processes have low reaction rates compared with the bimolecular reaction with O$_2$ (typically 10$^{-31}$ cm$^6$ and 10$^{-17}$ cm$^3$ s$^-1$) and the net formation rate is lower by a factor 10$^4$ and 10-100 for termolecular and radiative processes, respectively. According to \cite{gou12}, both oxygen and magnesium addition processes to grow silicate clusters are down-hill and no activation barrier is considered for the rates.

Ionisation of atoms in the ejecta occurs via collision with Compton electrons. The radioactive decay of \Ni, to \Co,~and \Fe~creates \grays~ that degrade to X rays and UV photons through collision with thermal electrons, thus inducing the creation of a population of Compton electrons in the gas. These fast electrons ionise atoms and destroy chemical species in the ejecta. 
The time-dependent destruction rate by Compton electrons $k_C$ for species $i$ in s$^{-1}$ is calculated using Eq.~4 of \citet{cher09}. The rate is re-scaled according to the amount of \Ni~produced by the explosion of the various progenitors, and following \citet{cher09}, the rate values are converted to a Arrhenius temperature-dependent form whose parameters are listed in table \ref{tbl-A2} in the Appendix. The interaction of the Compton electrons with molecules leads to their dissociation, ionisation and fragmentation into ionic products. The branching ratios for the different processes depend on $W_i$, the mean energy per ion-pair for a given species. Available values of $W_i$ for molecules that form in the ejecta are listed in table \ref{tbl-A2}. When data are not available, we simply assume values similar to those for O for elements and CO for molecules. The impact on molecules and dust clusters of the ultra violet (UV) radiation field resulting from the degradation of $\gamma$-rays was assessed by \citet{cher09}, who found that the destruction of molecules and dust precursors by this UV radiation field was not important compared with destruction by Compton electrons. We thus ignore UV radiation for the rest of the present study. 

\section{RESULTS}
To better understand the chemical composition of several post-explosion ejecta reflecting the evolution and nucleosynthesis of massive stars, we model the formation of molecules and dust clusters in the ejecta of four SNe with progenitors of mass 12, 15, 19, and 25 \Ms. The mass of \Ni\ is either 0.075 or 0.01 \Ms. The 15 \Ms\ progenitor with M(\Ni)$= 0.075$ \Ms\ is chosen as the "standard case", for which results on molecules and dust are presented in \$ \ref{15}. The impact of varying the \Ni\ mass is studied in \$ \ref{ni}. More massive progenitors are considered in \$ \ref{1925}, while results for a low-mass progenitor with a small \Ni\ mass are given in \$ \ref{12}. Finally, results on elements are shown in \$ \ref{el} and various dust formation models in SN are compared in \$ \ref{com}. 

\subsection{15 \Ms\ Progenitor} 
\label{15}
We present the masses of molecules, dust clusters and elements as a function of time after explosion (in days) for the "standard case". The chemistry is followed from day 100 until day 1500, a time span that covers the initial formation of molecules at early times until the dust cluster synthesis is fully completed some 4 yr after outburst. 

\subsubsection{Molecules}
\label{mol15}
We find that the zones of the He-core are efficient at forming large amounts of molecules. Because the ejecta is assumed to be hydrogen-free, the number of chemical species formed is limited; this poor chemistry typical of SN ejecta is well exemplified by the detection of only two molecules, CO and SiO, in several SNe \citep{dan88, roch91,ko05}, and CO in SN remnants \citep{rho09, rho12}.  

\begin{figure}
\figurenum{2}
\epsscale{1}
\plotone{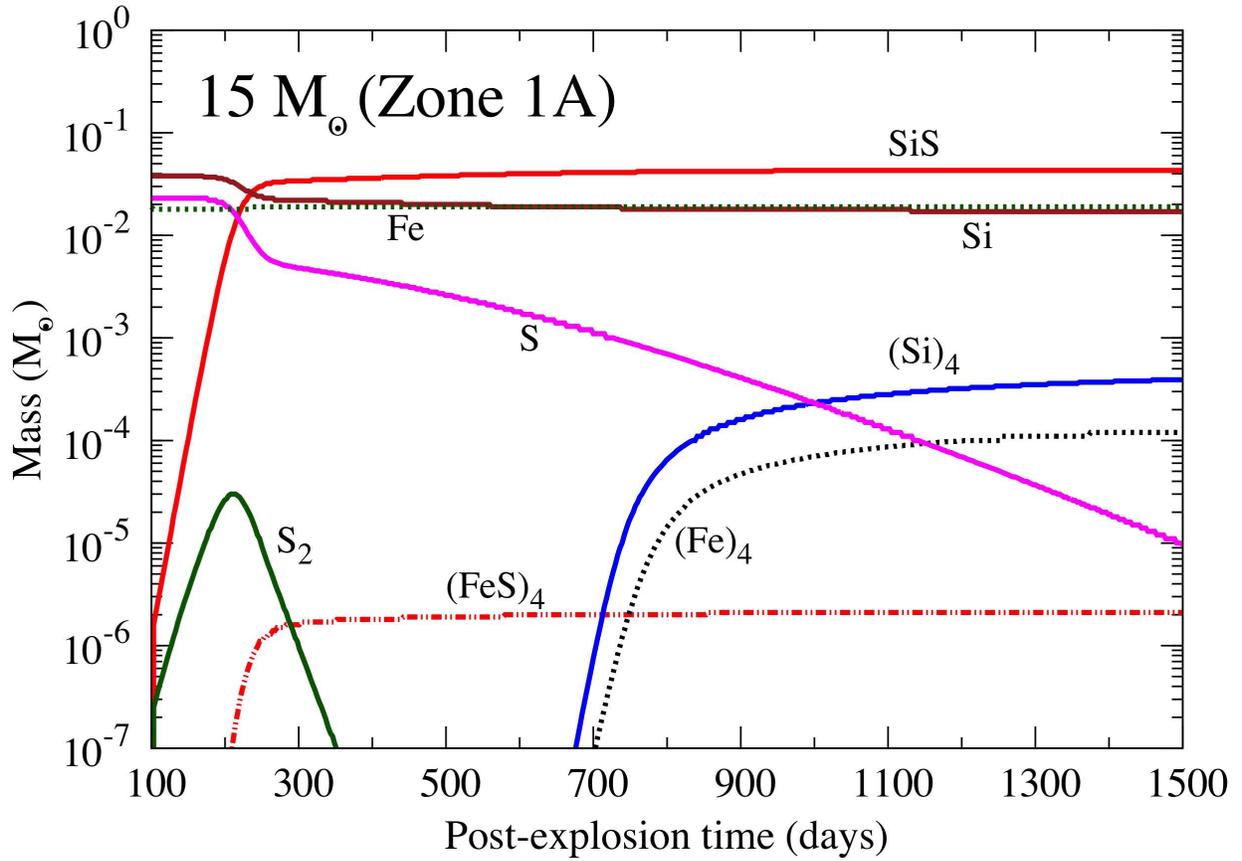}
\caption{Molecules formed in the innermost zone, zone 1A, of the 15 \Ms~ejecta. SiS is the prevalent species and depletes both Si and S atoms. Small masses of pure Fe clusters and iron sulphide, FeS, clusters also form.\label{fig2} }
\end{figure}

The innermost layer, zone 1A, is rich in iron, silicon and sulphur, where the iron results from the decay of \Ni~and \Co. The oxygen content of the zone is very low and precludes the formation of O-bearing species, metal oxides and silicates. Zone 1A quickly converts most of the atomic sulphur and half of the atomic silicon mass into silicon sulphide, SiS, as illustrated in Figure \ref{fig2}. The SiS mass rapidly increases after $\sim 200$ days to reach $4.3 \times 10^{-2}$ \Ms~1500 days post-explosion. The main formation processes for SiS is the radiative association reaction 
\begin{equation}
\label{SiS1}
S + Si \rightarrow SiS + h\nu,
\end{equation}
and the reaction 
\begin{equation}
\label{SiS2}
S_2 + Si \rightarrow SiS + S.
\end{equation}

The latter process is not well characterised with no measured reaction rate, and was included based on a process involving the reaction of atomic C with disulphur, for which a rate has been estimated at 300K. Owing to the isovalence of carbon and silicon, a similar rate was adopted for Reaction \ref{SiS2}. We tested the importance of Reaction \ref{SiS2} in the formation of SiS by removing it from the chemical  network. SiS was then mainly formed by Reaction \ref{SiS1} with similar efficiency and masses. Therefore, we conclude that the innermost zone of the He-core overwhelmingly produces SiS due to the large S and Si content of the zone. 

\begin{figure}
\figurenum{3}
\epsscale{1.}
\plotone{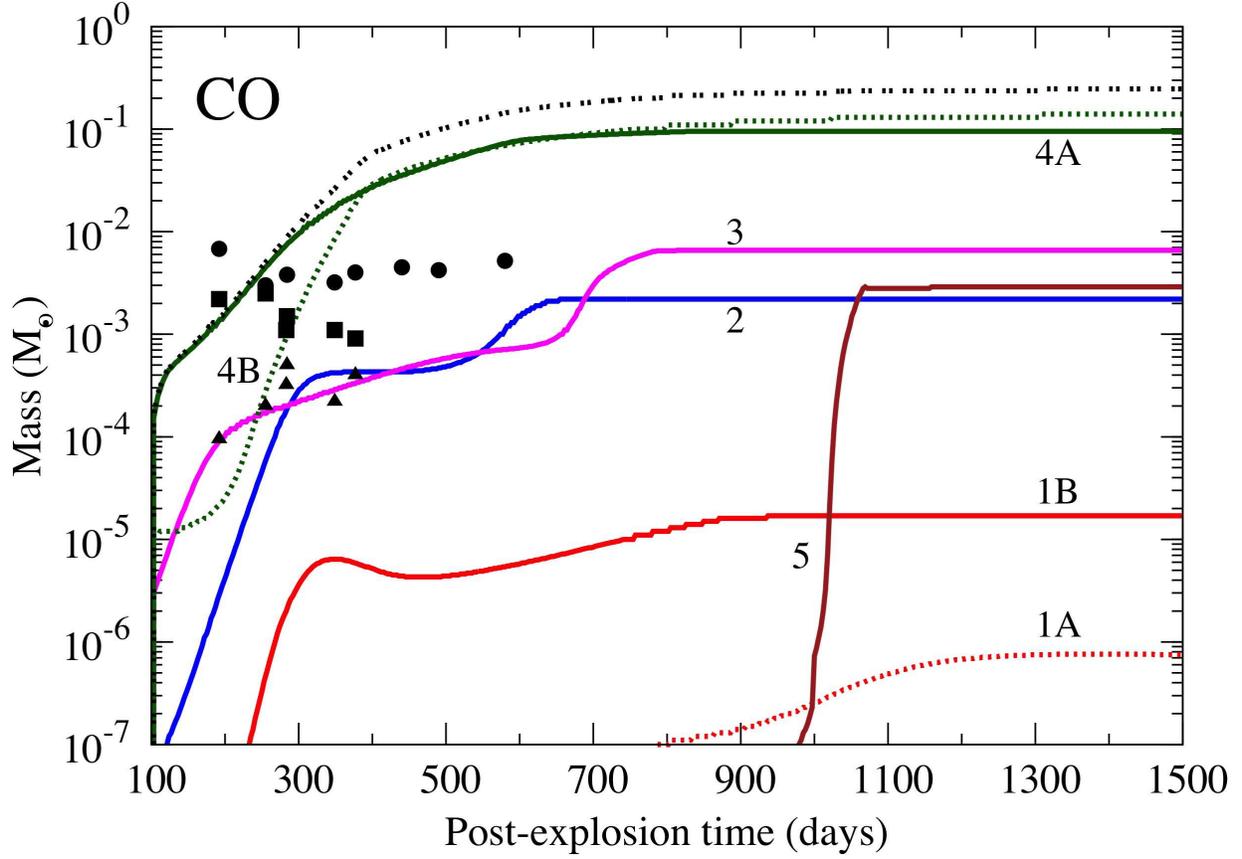}
\caption{Evolution of CO masses with post-explosion time for the 15 \Ms~progenitor as a function of ejecta zones (see Table 1 for zone labelling). The dotted-grey line represents the mass summed over all zones. CO is prevalently produced by zones 4A and 4B, followed by zones 2 and 3. The CO mass reaches large values ($\sim 10^{-1}$ \Ms) some 4 yr after explosion. CO masses derived from observations for SN1987A are also shown as symbols: LTE (triangle) and non-LTE (square) assumption \citep{liu92}, and thermal (round) assumption \citep{liu95}.\label{fig3}}
\end{figure}

Carbon monoxide, CO, was the first molecule detected in SN1987A. The fundamental band at 4.65 \mic\ was observed between day 135 and day 260 \citep{dan88}, while the CO first overtone transition at 2.29 \mic\ was detected at day 100 after the explosion \citep{spi88}. CO was later been detected in several other SNe \citep{cher11}. Once formed, CO can withstand harsh conditions in the ejecta because of its strong chemical bond. Depending on the C/O ratio characterising each zone, CO formation limits the amount of left over atomic oxygen or carbon in the gas-phase, and thus controls the chemistry of the gaseous and solid components of the gas. In the present case, most of the He-core zones have C/O ratios less than one, except for the outer mass zone, zone 5 (see table \ref{tbl-1}). The evolution of CO mass with post-explosion time for the He-core zones is shown in Figure  \ref{fig3}. CO masses derived from available observational data for SN1987A are also plotted for early times. In zones 4A, 4B, 2 and 3, CO forms as early as 200 days and reaches masses ranging from  $10^{-4}$ to $10^{-2}$ \Ms. The prevalent formation processes between 100 and 200 days are 

\begin{equation}
\label{CO1}
O + C_2 \rightarrow CO + C,
\end{equation}
and 
\begin{equation}
\label{CO2}
C + O \rightarrow CO + h\nu.
\end{equation}
The formation of C$_2$ chains via radiative association reactions starts early on, but owing to the large atomic oxygen content, C$_2$ is quickly converted to CO following Reaction \ref{CO1}. The formation of CO via oxidation of carbon chains prevails at early times, while Reaction \ref{CO2} contributes to the growth of CO mass after day 300. The final CO mass summed over all zones at day 1500 is $\sim 2\times 10^{-1}$ \Ms, much larger than the masses derived from IR data before day 600 in SN1987A \citep{liu95}. These large amounts of CO primarily form in zones 4A and 4B, and do not trace efficient carbon dust formation in these two zones. These zones indeed form little or no carbon dust because carbon chains are quickly destroyed by oxidation reactions similar to Reaction \ref{CO1} to form CO, thus impeding their growth into larger carbon clusters. 

The oxygen-rich component of the He-core includes zones 1B, 2, 3, 4A, and 4B, extending from 1.88 \Ms~to 3.04 \Ms~(see table \ref{tbl-1}). Oxygen-bearing molecules are expected to form there but the zones are rich in inert gas, namely Ar (Zone 1B) and Ne (Zones 2 and 3). Ar and Ne atoms are ionised by Compton electrons, and the ions (\arp~and \nep) are destroyed by both the recombination to their inert parents and the shrinkage in Compton electrons with time owing to the decreasing mass of \Ni. \arp~and \nep are detrimental to the formation and survival of molecules in these zones as they quickly destroy molecules. The mass of silicon oxide, SiO, formed in the various zones as a function of post-explosion time is shown in Figure \ref{fig4}. Superimposed are the SiO masses derived from IR observational data for several SNe. The mass follows a rapid increase at day 200 in zones 1B, 2, and 3, while the formation of SiO is delayed to 400 days in zones 4A and 4B. The prevalent formation process for SiO in zones 1B, 2, and 3 are the reactions 

\begin{equation}
\label{SiO1}
Si + O \rightarrow SiO + h\nu,
\end{equation}
\begin{equation}
\label{SiO2}
Si + O_2 \rightarrow SiO + O,
\end{equation}
and 
\begin{equation}
\label{SiO3}
Si + CO \rightarrow SiO + C.
\end{equation}
The prevalent destruction processes are 

\begin{equation}
\label{SiO4}
SiO + Ar^+ \rightarrow Si + O + Ar^+,
\end{equation}
for zone 1B,
\begin{equation}
\label{SiO5}
SiO + Ne^+ \rightarrow Si + O + Ne^+,
\end{equation}
for zones 2 and 3, and
\begin{equation}
\label{SiO6}
SiO + O \rightarrow Si + O_2.
\end{equation}
for both zones. Upon formation, SiO is destroyed by \arp~and \nep~following Reactions \ref{SiO4} and \ref{SiO5}. The SiO mass for all zones shows a gradual and strong decrease, going from $\sim 10^{-2}$ \Ms~at 200 days to $\sim 10^{-6}$ \Ms~at 1500 days. Such a decline is also shown by the SiO masses derived from observations, for example, in SN2004et \citep{ko09}. In this object, the SiO transition was detected at various periods and showed a gradual fading with time that is coupled to the evidence for dust synthesis in the ejecta. The destruction of SiO before 400 days results from thermal fragmentation and the destruction by Ar$^+$ and Ne$^+$ ions. At later times, SiO is depleted into silica and silicate clusters, as we will see in the next section, and acts as a direct dust synthesis tracer in the ejecta. 

Apart from SiO, the formation of dioxygen, O$_2$, and monosulphide, SO, prevails in the O-rich zones of the ejecta, as illustrated in Figure \ref{fig5}. Most of O$_2$ and SO molecules form in both zones 2, 3, and 4A. Their mass variation shows a similar trend with time, ranging from negligible masses before day 600 and reaching high mass values after day 600. Dioxygen efficiently forms at early time from the radiative association reaction 
\begin{equation}
\label{O21}
O + O \rightarrow O_2 + h\nu,
\end{equation}
 but is quickly depleted in the formation of SiO, and, to a minor extent CO, following Reaction \ref{SiO2} and the reaction
\begin{equation}
\label{O22}
C + O_2 \rightarrow CO + O.
\end{equation}

At later times, the formation of AlO also contributes to the destruction of O$_2$ via the reaction
\begin{equation}
\label{AlO1}
Al + O_2 \rightarrow AlO + O,
\end{equation}
 
while the reverse of Reaction \ref{AlO1} contributes to the reformation of O$_2$. The gradual depletion of SiO in silicate clusters allows the O$_2$ mass to grow after day 600 to reach the large value of $\sim 4 \times 10^{-1}$ at day 1500. The SO mass follows a trend similar to that of O$_2$ because the SO formation is directly coupled to the formation of O$_2$ by the reaction 
\begin{equation}
\label{SO1}
S + O_2 \rightarrow SO + O,
\end{equation}

and the SO destruction follows the reverse of Reaction \ref{SO1}. The final SO mass is large and amounts to $2\times 10^{-2}$ \Ms~at day 1500. 

We see that all chemistries responsible for the production of molecules are entangled, and the final molecular component of the He-core includes five molecules, namely CO, O$_2$, SiS, SO, and SiO. Aluminium oxide, AlO, is not included in the molecular component because it will quickly be depleted in (AlO)$_2$ and alumina dust clusters (see \S \ref{dust15}). We consider AlO as a "dust cluster" rather than a gas-phase molecule). The first four molecules form in the ejecta, participate in the ejecta chemistry prior to day 1500, and are ejected with large masses to later stages of the SN evolution, e.g., in the SN remnant phase. The SiO molecule, on the other hand, forms efficiently but is quickly depleted into the production of dust clusters after day 300, and as such, enters the SN remnant phase with a much smaller mass than the other species. The total mass of the molecular component of the SN ejecta is high, and summarised as a function of zoning in table \ref{tbl-4}. The final ejecta mass fraction residing in molecules at day 1500 amounts to $\sim 30$ \% of the ejected mass for the 15 \Ms~progenitor. 

\subsubsection{Dust}
\label{dust15}

As discussed in \S \ref{chem}, the description of the dust synthesis is based on the formation of large molecular clusters entering the nucleation phase of the dust grains. The nucleation phase involves the chemical kinetic description of the formation of these clusters from the gas phase. For silicates, our larger clusters are dimers of forsterite (Mg$_4$Si$_2$O$_8$), while for carbon dust (possibly solid C$_{60}$), we model the formation of the first carbon ring C$_{10}$. As for alumina, we are currently working on a chemical scheme to model alumina cluster formation based on cluster structures and the calculation of chemical rates but it is too premature to include such a scheme in our model. The most stable structure of \al\ is kite-shape \citep{arch99}, and the formation of molecular \al\ probably involves the dimerisation of AlO and the possible addition of one oxygen atoms via a bimolecular reaction with a O-bearing species. We can then safely assume that AlO molecules are precursors to alumina via the formation of (AlO)$_2$ and that the AlO mass indicates an upper limit on the \al\ mass that can form in the ejecta. The condensation phase involves the coalescence of these clusters with each other, combined with surface growth if gas-phase growing agents are available. The condensation phase is not considered in the present study and the calculated masses of clusters thus represent an upper limit on the total mass of dust.

The modelled masses of dust clusters over all zones are illustrated in Figure \ref{fig6} for the standard case. As discussed above, the condensation phase is not modelled and the cluster mass curves then flatten to their upper limit values once nucleation has taken place. As seen in Figure \ref{fig6}, there exist various events of cluster formation in the ejecta according to the chemical type of the dust and the zones in which the clusters form. The FeS clusters are first to form at day 250 in the innermost zone, zone 1A, followed by silica and forsterite clusters in zones 1B and 2 at day 350. Aluminium monoxide AlO forms after day 600 in zones 2 and 3. Most pure metal clusters form after day 700 and include Mg and Si in zones 2 and 3, and iron in zone 1A. Finally carbon and silicon carbide clusters are synthesised in the outermost zone, zone 5, at late times ($\sim$ 1050 days). 

The timing of dust production highly depends on the local chemistry characterising the zones, as exemplified by the formation of silicates and carbon clusters. The forsterite dimer mass curve obtained for the various zones of the ejecta is shown in Figure \ref{fig7}. Forsterite first nucleates as early as 300 days in zone 1B, and gradually grows to reach its maximum value at $\sim$ 900 days in this zone. These two nucleation phases are seen as two depletion events in the SiO mass curve for zone 1B (see Figure \ref{fig4}), and correspond to the formation of the O$_2$ molecule at 300 days and 900 days in zone 1B (see Figure \ref{fig5}). The subsequent oxygen addition to silicate intermediates grows forsterite dimers. In zone 2, forsterite forms at $\sim$ 600 days from the depletion of SiO (see the drop in Figure \ref{fig4} for zone 2) and the net formation of O$_2$ molecules that permits the final growth to forsterite dimers. The formation of forsterite dimers is less effective in the other three zones of the ejecta and occurs at late epochs. Therefore, the gradual growth of the forsterite total mass as shown in Figure \ref{fig6} results from the chemistry of SiO formation and the growth of silicate clusters following the two-step mechanism of oxygen and magnesium addition proposed by \citet{gou12}.

The scenario for the growth of carbon small clusters is quite different. First of all, our results highlight the fact that carbon chains grow in significant amounts in the only carbon-rich zone of the ejecta, zone 5. Where oxygen overcomes carbon, e.g., in zone 4A and 4B, any carbon chain is destroyed by O attack to form CO. Zone 5 is helium-rich and He atoms are ionised by Compton electrons. The produced ions are destroyed by recombination to He and by the decrease of Compton electrons with time. \hep~is detrimental to the formation and survival of molecules in zone 5 \citep{lepp90,cher09, cher10a} as the ion quickly destroys molecules following reactions such as
\begin{equation}
\label{He1}
He^+ + C_n \rightarrow C_{n-1} + C^+ + He.
\end{equation}
Once the \hep~ion mass becomes negligible after day 1000, molecules like CO, SiC, C$_2$ and CS quickly form. The growth of carbon chains is then efficient via radiative association reactions of the type 
\begin{equation}
\label{C1}
C + C_n \rightarrow C_{n+1} + h\nu.
\end{equation}
As the zone is C-rich, the low oxygen content hampers the destruction of carbon chains via reaction such as Reaction \ref{O22}, which grow until the closure of the first ring, C$_{10}$. The formation of carbon clusters therefore strongly depends on the \hep~content of the outer zone and the time at which the \hep~abundance significantly decreases. Because of the very large initial He mass, the vanishing of \hep~in the zone is delayed to $\sim$ day 1050, resulting in a late formation of carbon and silicon carbide clusters. This late synthesis contributes to the time-sequence of dust cluster production observed in Figure \ref{fig6}. 

The gradual increase with time in cluster masses results in upper limits on dust mass ranging from 10$^{-5}$ \Ms~at $\sim$ 300 days to $\sim 4 \times 10^{-2}$ \Ms~more than 4 yr after explosion. This range of dust masses perfectly agrees with the values derived from observational data. At early times ($200 < t < 600$ days), the mass derived from IR observations in several Type II-P SNe are small with typical values between $10^{-5}$ \Ms~and $10^{-3}$ \Ms. These values correspond to our modelled cluster masses shown in Figure \ref{fig6} for this time span. However, much larger dust masses are derived in SN remnants from submm data. In Cas A, $\sim 0.08$ \Ms~of dust is inferred from the Herschel data \citep{bar10}. In SN1987A, between 0.4 and 0.7 \Ms~ of dust is inferred to reproduce the Herschel fluxes \citep{mat11}, while $0.1-0.2$ \Ms~of dust is found in the filamentary structures of the Crab Nebula by \citet{gom12} with Herschel. These masses obviously result from fitting the spectral energy distribution of the objects at IR and submm wavelengths, implying that an initial dust composition is assumed. In the case of Cas A and the Crab Nebula, two types of dust are considered separately, namely amorphous carbon (AC) and silicates, while for SN1987A, large iron spheres are also included to obtain a satisfactory fit of the flux data. Notwithstanding the presence of cool dust in SNRs in amounts larger than what is observed at early times in SN ejecta at IR wavelengths, the derived dust masses and chemical compositions are somewhat uncertain, and strongly depend on the physical and chemical parameters used in SN ejecta and remnants (e.g. the dust chemical composition, grain size distribution and temperatures). The chemical composition of the dust is illustrated for all progenitor masses in Figure \ref{fig8}, where we assume that all clusters are spontaneously depleted in dust grains with 100 \% efficiency at all times. For our standard case, the final dust composition and mass fraction at day 1500 consist of 60 \% amorphous carbon , 20 \% alumina, 15 \% forsterite, and a few \% pure metal clusters, namely, Si, Mg, Fe, silica, and silicon carbide clusters. This dust composition is different from that assumed in analysis of IR and submm data and reflects the chemically-controlled nucleation of dust clusters. The formation sequence and the gradual growth of dust clusters with time in the ejecta is a consequence of the ejecta chemistry that produces molecules combined with the nucleation processes of the various clusters. Hence, this gradual growth of clusters represents a genuine explanation for the existing discrepancy between the small dust masses found in SNe and the larger dust masses inferred in SNRs. 

\subsection{Impact of the \Ni~Mass}
\label{ni}

The \Ni~mass produced by SNe can be derived from the variation of the optical light curves and H$_{\alpha}$ luminosities in the nebula phase \citep{elmb03}. From direct identification or by comparison with explosion models, a mass for the supergiant progenitors can be inferred. Table \ref{tbl-5} lists some Type II-P SNe, with estimated \Ni~masses and progenitor mass range.  Most of the SNe have progenitor masses between 10 and 15 \Ms~and typical \Ni~masses of 0.01-0.02 \Ms, except for SN1987A and SN1999em, which have much larger values for both progenitor and \Ni~mass. This dichotomy reflects the trends derived by \citet{ham03}, that more massive SNe produce more energetic explosions, and SNe with greater energies produce larger \Ni~masses. To account for the low \Ni-mass SNe, we study the impact of the \Ni~mass on the ejecta chemistry of our standard 15 \Ms~progenitor, while results for a low-mass progenitor (12 \Ms) with a low \Ni~content (0.01 \Ms) are presented in \S \ref{1925}. 

The primary impact of a smaller \Ni~mass in the ejecta is to reduce the number of Compton electrons resulting from the degradation of a smaller amount of \grays. Therefore fewer ions such as \arp, \nep, and \hep~are produced, enhancing the survival of molecules and clusters. The mass evolution of forsterite clusters and carbon rings versus post-explosion time are illustrated in Figure \ref{fig9} for the two \Ni\ mass values, 0.01 \Ms\ and 0.075 \Ms. The destruction of molecules from which clusters form (e.g., SiO and C$_2$) is not as severe for the low \Ni~mass case as it is for the standard case, because of the lower \nep~and \hep~ejecta content. Therefore, the formation of all clusters proceeds at early times and at large gas densities, resulting in a larger molecular component and dust cluster formation efficiency in the ejecta, as seen from table \ref{tbl-4}. Most importanty, the impact of reducing the \Ni~mass anticipates the formation of forsterite clusters as early as at 250 days in zone 2. These clusters may not coalesce readily into silicate grains because the gas temperature is still high ($\sim 2300$ K) at day 300, but the efficient formation of silicate clusters results in changing final dust cluster compositions at late time $-$ see table \ref{tbl-4} and Figure \ref{fig8}. The dust budget now includes 45.0 \% forsterite, 41.4 \% carbon, and 13.6 \% alumina, as opposed to 60 \% carbon, 20 \% alumina, 15 \% forsterite, and a few \% pure metal clusters for the standard case of \S \ref{dust15}. 

\subsection{19 \Ms\ \& 25 \Ms\ Progenitors}
\label{1925}
We now present the results for the ejecta chemistry associated with a 19 \Ms~progenitor. The initial composition of the ejecta is given in table \ref{tbl-1}, and the chemistry for all zones is identical to that considered for the standard model. All results for molecule and dust masses are summarised in table \ref{tbl-6}. The masses of CO and SiO formed over the time-span 100-1500 days are shown in Figure \ref{fig10}. The evolution of both molecules with post-explosion time resembles that of the 15 \Ms~case. The formation of CO commonly occurs in zones 3 and 4, which correspond to zones 4A and 4B of the 15 \Ms~progenitor. Theses zones are not efficient at forming dust, as seen in table \ref{tbl-6}. As for the 15 \Ms~case, the conclusion that CO does not trace carbon dust formation in the ejecta holds for the 19 \Ms~progenitor. Likewise, the variation of SiO masses with time shows similar trends as for the standard model, the only exception being zone 2. Indeed, the destruction of SiO prevails at early times owing to the high Ne content of the zone, as seen in table \ref{tbl-1}, that results in high \nep~masses at early times. The chemical trends for the formation and destruction of O$_2$ and SO are also akin to those already described in \S \ref{mol15}. More generally, the total mass of molecules produced by the 19 \Ms~progenitor is higher by a factor of $\sim$2 compared with the 15 \Ms~case, but the efficiencies of forming molecules for the two cases are similar. About 30 \% of the material ejected by a SN with a progenitor mass of $15-19$ \Ms~is in molecular form. 

In Figure \ref{fig11}, we show cluster mass evolution as a function of time for the 19 \Ms~progenitor. The FeS clusters are the first to form in Zone 1A with a small mass compared with pure iron clusters (see table \ref{tbl-6}). Furthermore, the FeS mass is smaller than that derived by Cherchneff \& Dwek (2010) for the 20 \Ms\ progenitor with zero metallicty. This discrepancy arises from the larger Fe/S yields characterising the innermost zones of the primeval 20 \Ms\ progenitor. Forsterite clusters experience two phases of growth. The first phase between 200 and 500 days is characterised by a forsterite mass reaching a value $\sim 5 \times 10^{-4}$ \Ms~at 300 days. Zone 1B is responsible for this early growth event owing to the first production event of SiO at day 200 seen in Figure \ref{fig10}. A second forsterite growth event occurs around day 700, corresponding to the peak in SiO formation in zone 2 at this time. The composition of the grains is shown in Figure \ref{fig8}. The prevalent dust formed is alumina, followed by forsterite, carbon, and finally some Mg dust. Despite similarities between the chemical processes at play in the formation and destruction of molecules and dust in the ejecta, and similar upper limit values on the final dust mass at day 1500, the variation in the dust composition between the 15 \Ms~and the 19 \Ms~progenitors reflects the initial chemical composition of the ejecta given by the explosion nucleosynthesis models. 

We now consider the ejecta of a 25 \Ms~progenitor with a high \Ni~mass (0.075 \Ms), a surrogate for the explosion of a massive star like the progenitor of SN2004et, or the explosion of population II supergiant stars at high redshift, metallicity not with standing. The cluster masses versus post-explosion time are shown in Figure \ref{fig11} and the molecule and cluster masses as a function of ejecta zoning are listed in table \ref{tbl-7}. The synthesis of clusters starts at day 200 with a rapid increase in the forsterite mass that reaches $6 \times 10^{-3}$ \Ms\ at day 400 and $6 \times 10^{-2}$ \Ms\ at 700 days. As seen from table \ref{tbl-1}, a large oxygen and silicon content characterises the oxygen core zones 2 and 3 of the 25 \Ms\ progenitor, and triggers  an efficient silicate cluster synthesis. The large aluminium and oxygen yields characterising zone 3 guarantee the formation of a large mass of AlO, the molecular precursor to \al. However, the large Ne yield of the zone results in a large \nep\ abundance that delays the formation of AlO to day 750. Table \ref{tbl-1} also indicates a large fraction of carbon compared with oxygen (C/O $\sim 36$) accompanied by a large yield of helium in zone 5. Such a composition results in the delayed formation of carbon and silicon carbide clusters in this zone after day 1500 owing to the large fraction of \hep~in the zone. As illustrated in Figure \ref{fig8}, the 25 \Ms\ progenitor primary forms alumina and forsterite clusters, while carbon dust is a very minor component of the condensates formed by these massive SNe. 

To conclude, SNe with large progenitor masses tend to form dust with efficiencies similar to that of the standard 15 \Ms\ progenitor, but are more efficient at forming molecules. The molecular component of the ejecta can be as large as $\sim$ 50 \% of the total ejected mass. The larger the progenitor mass, the later carbon clusters form owing to the large \hep\ content in the outermost ejecta zone. At late time, the decrease in gas number densities may hamper the efficient condensation of carbon chains in AC dust. In the end, these massive progenitors produce a small mass of carbon clusters that may not totally transform into dust, but should chiefly synthesise O-rich condensates (e.g., silicates and alumina) in their ejecta.

\subsection{12 \Ms\ Progenitor with a Low \Ni\ Mass}
\label{12}

As shown from table \ref{tbl-5}, low-mass progenitors tend to produce low \Ni~mass in contrast with large-mass progenitors. We thus model the ejecta of a 12 \Ms~progenitor with a small \Ni~mass (0.01 \Ms), that can be regarded as a template for low-energy, faint SNe. Another SN environment originating from a low-mass progenitor includes the Crab Nebula, a SN remnant resulting from the explosion of a supergiant with mass $\sim$ 10 \Ms\ \citep{dav85, alp08}. The molecule and cluster masses as a function of ejecta zoning for the 12 \Ms~progenitor are given in table \ref{tbl-8}. At large, results akin to those for other progenitor masses are obtained, i.e., the ejecta produces a large fraction of molecules ($\sim 20$ \% by mass). The ejecta zone most efficient at producing species including CO and O$_2$ corresponds to the O-and C-rich outer zone of the oxygen core (labelled zone 3) characterised by a C/O ratio with a typical value of 0.3. For the 12 \Ms\ progenitor, the innermost zone, zone 1A, is also very efficient at producing SiS because of the large initial Si and S yields and the low Fe yield, as seen in table \ref{tbl-1}. Essentially all atomic S gets trapped in SiS in this zone.  

As for dust, cluster masses versus post-explosion time are shown in Figure \ref{fig12}. As discussed in \S \ref{ni}, the formation of forsterite clusters occurs as early as 250 days post-explosion owing to the low \Ni\ ejecta content. However, the gas temperature at day 250 is too high ($\sim$ 2500 K) to permit the coalescence of forsterite clusters in silicate dust. This process will take place once the gas temperature reaches $\sim$ 1500 K around day 400. However, for this progenitor, the onset of AlO formation also occurs around day 250. Alumina, \al, being more refractory than silicates,  may then precede the formation of forsterite between 250 and 300 days, leading to the early formation of \al\ dust in the ejecta, followed by a forsterite dust formation event. As for the 15 \Ms\ low \Ni\ case, the synthesis of carbon chains and rings is also shifted to an earlier epoch, $\sim$ 800 days, implying that AC grains will form more efficiently in low-mass SN progenitors because of the larger gas densities. A 12 \Ms\ model should thus lead to a dust formation time$-$sequence of \al\, silicate, AC, where in the end, carbon represents $\sim 68$ \%, silicate $\sim$ 28 \%, and alumina  $\sim$ 6 \% of the total dust mass, respectively, as seen from Figure \ref{fig8}. 

A dust formation event before 200 days, possibly involving alumina as a first condensate, should then characterise SN progenitors with masses smaller than 12 \Ms\ and low \Ni\ masses. Such an early dust formation episode is observed in some SNe with low-mass progenitors, e.g., SN2011ht \citep{mau12}, and is often ascribed to the interaction of the explosion blast wave with the dense progenitor wind at early epochs. A dense shell conducive to dust condensation forms in the post-shock region, resulting in a dust production episode. The present results indicate that the early formation of dust may also be due to a low \Ni\ mass in the ejecta. Hence, the observed early condensation episode in faint SNe may arise from a combination of both scenarios.   

\subsection{Elements}
\label{el}

Sections \S \ref{mol15} and \S \ref{dust15} highlighted the importance of the molecular and dust components of SN ejecta, that amount to 30\%$-$50 \% of the ejecta mass depending on the progenitor mass. The rest of the ejecta is in the form of atomic elements, either neutrals or ions. The masses of elements as a function of post-explosion time are shown in Figure \ref{fig13} for the standard case. The top figure illustrates the masses summed over all ejecta zones versus time, while the bottom figure focuses on atomic Si mass variation versus time for each zone. Until day \env\ 800, most elements including O, Mg, Fe, and Al show almost constant masses in the ejecta. After day 800, O and Al masses decrease owing to the formation of AlO in the O-rich core of the ejecta. The mass of atomic carbon shows a small decrease after day 300 due to the formation of CO in most of the zones, and a sharper decline after day 1050 resulting from the formation of carbon chains and rings in zone 5. The overall atomic silicon mass slightly decreases over time until it reaches a constant mass at day \env\ 700. This mass variation primarily reflects zone 1A where the large Si content is depleted at early epochs in the formation of SiS (see Figure \ref{fig2}). The Si dotted line in the top figure depicts the Si mass summed over all zones except for zone 1A. These zones include the oxygen-rich core of the ejecta (Zones 1B, 2, and 3) where most of the silica and silicate clusters form. The summed Si mass shows a sharp decline between \env\ 200 and 800 days that reflects the depletion of Si in the formation of SiO, silica and silicates in zones 1B, 2 and 3, between day 200 and day 700. 

As first shown by \citet{luc89} for SN1987A, the fluxes of the [OI] 6300 \AA\ and MgI] 4571 \AA\ emission lines faded with time, with a sharper decline at day $\sim$ 530, indicative of the onset of dust formation. The [SiI] emission line at 1.6445 \mic\ showed a markedly stronger fading relative to the continuum, that pinpointed either a decrease in the Si abundance owing to dust condensation or temperature effect induced by strong cooling in the Si zone. A similar fading of the Mg and O line fluxes from day 500 until day 800 observed by Lucy et al. indicates that extinction induced by dust condensation around day 500 was responsible for the fading. This interpretation is supported by the present calculations, as both elements show a time-independent mass evolution over this time span. On the other hand, the sharper fading observed in the Si emission line flux probably ensues from the combined effects of extinction and Si depletion in SiO and dust grains, as illustrated in Figure \ref{fig13}. 

\subsection{Comparison with Existing Studies}
\label{com}

Several studies have tackled the modelling of dust formation in Type II-P SNe. The first attempt to model the synthesis of grains in SN1987A was carried out by \citet{koz89}. Later studies dealt with the formation of dust in Type II-P SNe locally \citep{bian07} and at high redshift \citep{tod01,noz03}. All these studies consider the formation of dust grains from the gas phase using classical nucleation theory (CNT). Some assume a fully-mixed ejecta \citep{tod01, bian07} while others consider stratified ejecta \citep{koz89,noz03,koz09}. A few studies consider the impact of the steady-state formation of CO and SiO from the gas phase, including the destruction of CO by Compton electrons, on the final carbon and silicate dust mass \citep{tod01, bian07}. This assumption gives rise to the formation of carbon dust in a fully mixed ejecta with a C/O ratio less than 1, a result that contradicts the findings of \citet{koz89}. 

All existing CNT-based models for the progenitor masses of interest in the present study are summarised in table \ref{tbl-9}, with the derived dust masses and the modelled dust condensation sequences over time. CNT-derived dust masses for solar metallicity ejecta have values higher by a factor of $\sim 10$ compared with the upper limits of dust masses derived in this study. This discrepancy follows from several assumptions. First, several models \citep{tod01,bian07} consider a fully-mixed ejecta. Such a scenario is not confirmed by explosion hydrodynamic models \citep{ham10}, and observations of SN remnants which point to the memory of nucleosynthesis layers within the remnant, as in Cas A \citep{isen12}. Because the dust mass is derived from the total elemental yields and chemistry is not properly considered, fully-mixed ejecta always produce larger dust quantities. Second, in unmixed models, CNT is applied to steady-state conditions that are usually not found in the dynamic environments characterising SN ejecta. Finally, all CNT-based models ignore the non-equilibrium chemistry related to the formation of molecules and dust clusters, and the specific physics of SN ejecta where radioactivity greatly impacts the gas-phase chemistry through Compton electron ionisation. 

These specificities also affect the dust condensation sequence, as seen in table \ref{tbl-9}. In CNT-based studies, the condensation sequence is derived assuming equilibrium  temperature and pressure as initial conditions. In the present study, the dust condensation sequences ensue from non-equilibrium chemical kinetics and thus depend on ejecta parameters such as the initial post-explosion elemental yields, the mass of \Ni\ produced, and the gas temperature and density. This fact is well illustrated by the 15, 19, and 25 \Ms\ progenitors, for which silicate clusters form before the molecule AlO, when alumina, \al, is supposed to be the first solid to condense in O-rich environments at thermodynamic equilibrium \citep{tie98}. Here, \sif\ production precedes that of \al\ because of the early destruction of AlO molecules by \nep\ ions in the gas. 

More generally, existing studies based on CNT overestimate the total dust mass formed in SN ejecta. Dust formation sequences assuming thermodynamic equilibrium are very commonly used as benchmarks in the modelling of dust synthesis in O-rich evolved stellar media, but should be avoided when modelling stellar outflows and ejecta, where dynamics and chemistry control the synthesis of condensates. 

\section{SUMMARY AND DISCUSSION}

We have investigated the synthesis of molecules and dust clusters in stratified ejecta of Type II-P SNe with solar metallicity. Our results highlight the following points

\begin{itemize}

\item Molecules including SiS, CO, O$_2$, and SO represent a large fraction of the gas-phase ejecta ($\sim$ 30\% by mass). Specifically, the CO mass increases from $10^{-4}$ at 100 days and gradually reaches $\sim$ 0.1 \Ms~1500 days post-explosion for all SN progenitors. This high CO mass forms in an ejecta zone where carbon dust does not condense, suggesting that most of the observed CO does not trace the carbon dust formation process in SN ejecta. 

\item The molecule SiO efficiently forms at an early epoch and is quickly converted into SiO dimers and silica and silicate clusters. The mass of SiO at day 1500 is $ \sim 10^{-6}$ \Ms\ or less. SiO is thus a direct tracer of dust formation in SN ejecta. 

\item The dust clusters form at different post-outburst epochs in various zones. Silicate clusters experience a delayed formation owing to the early destruction of O$_2$ and SO. The growth of silicate clusters via oxygen addition then occurs at $\sim$ day 500 for our standard case. Carbon chains and rings and silicon carbide clusters form in the outermost zone of the He-cores and at late times owing to the detrimental attack of He$^+$ on molecules. In more general terms, the dust mass gradually increases over time from $\sim 10^{-5}$ \Ms\ at 400 days to \env\ $0.03-0.09$ \Ms~after day 1500. This gradual synthesis of dust clusters over a time span of $\sim$ 4 yr provides a plausible explanation for the discrepancy observed between the dust masses derived from IR observations at early epochs and the larger masses of cool dust observed at submm wavelengths in SN remnants. 

\item The formation of dust clusters occurs according to a sequence of condensation events at various epochs. Low-mass progenitors experience anticipated dust formation events of essentially carbon dust with a minor silicate and alumina component, owing to the small masses of \Ni\ in the ejecta. High-mass progenitors primary form silicates and alumina dust, with a minor component of carbon dust. Compared with existing models of dust synthesis in Type II-P SNe based on Classical Nucleation Theory, our results indicate masses of synthesised dust that are smaller by a factor of $\sim 10$ and different dust condensation sequences and chemical compositions. 
\end{itemize}


The large fraction of the material expelled in a SN event is in molecular form (\env\ 20\%$-$50 \%) with a chemical composition including SiS, CO, O$_2$, and SO, depending on zoning. These four chemical species will pervade the late stages of SN evolution, i.e., the SN remnant not yet hit by the reverse shock. Evidence for molecules in SN remnants was presented by the detection of the first overtone transition of CO in the young remnant Cas A \citep{rho09}. The fundamental band at 4.56 \mic\ was subsequently observed with AKARI \citep{rho12}. As already proposed by \citet{cher11}, our results strongly suggest that a large fraction of cool CO (\env\ 0.1 \Ms) formed in the ejecta should pervade the remnant gas not yet shocked by the reverse shock and thus be detectable. The recent detection with ALMA of cool CO formed in the ejecta of SN1987A with a derived mass $\ge$ 0.1 \Ms\ confirms our present results \citep{kam13}. Most interesting are the large masses of SiS ($0.04-0.1$ \Ms) formed in the innermost zone of SN ejecta. Emission line analysis of SN remnants suggest that the remnant has retained some memory of the ejecta stratification due to nucleosynthesis, consistent with explosion models \citep{chev78,fes06,del10,isen12,gav12}. If so, SiS molecules should exist in the Cas A  remnant in sulphur, silicon, and calcium-rich fast moving knots, and possibly be detectable there at submm wavelengths. 

\begin{figure}
\figurenum{4}
\epsscale{1.}
\plotone{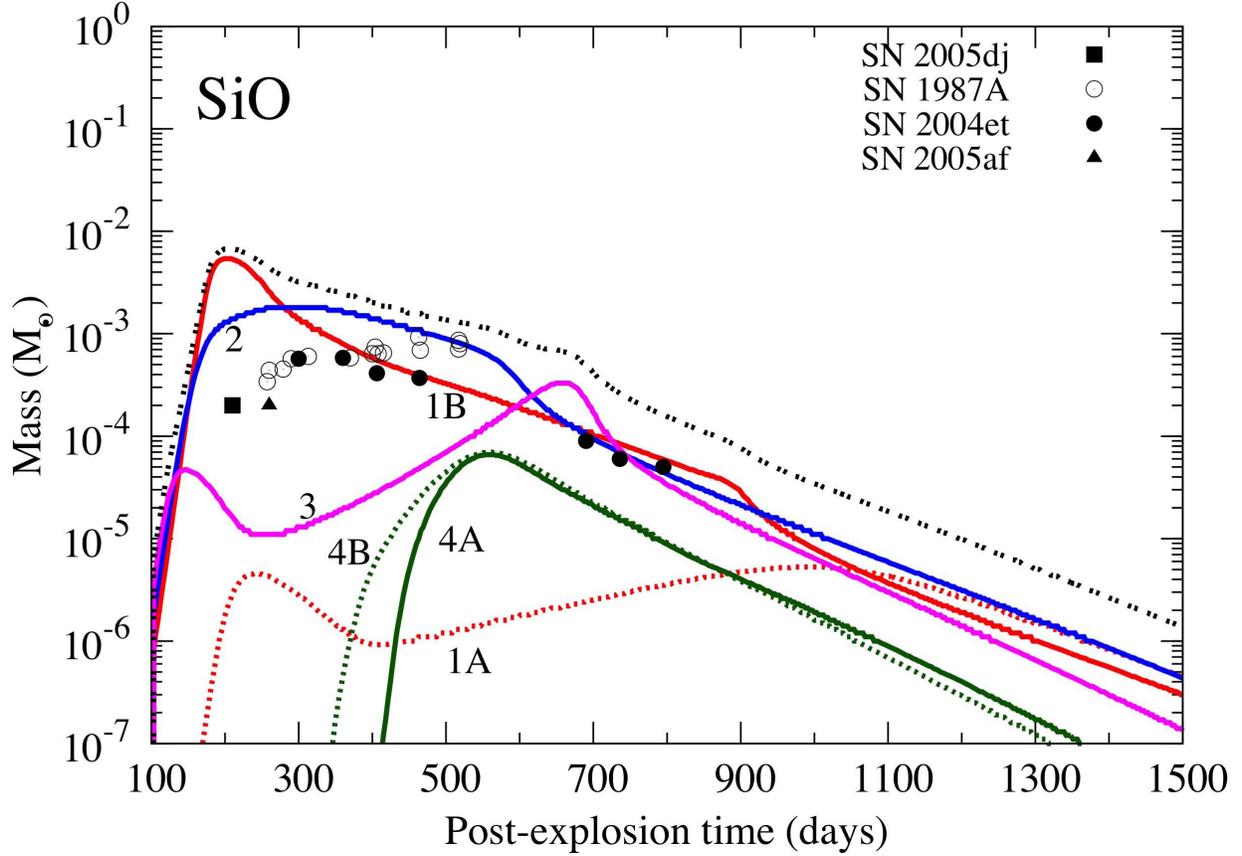}
\caption{Evolution of SiO masses with post-explosion time for the 15 \Ms~progenitor as a function of ejecta zones (see Table 1 for zone labelling). The dotted-grey line represents the mass summed over all zones. SiO formation prevails at early time in zones 1B and 2 when zone 3 also contributes at later epoch. The SiO mass shows a strong decrease that reflects the formation of silica and silicate clusters in the ejecta O-rich zones. The masses derived from the observations of several SNe are also shown as symbols. \label{fig4}}
\end{figure}

\begin{figure}
\figurenum{5}
\epsscale{1.8}
\plottwo{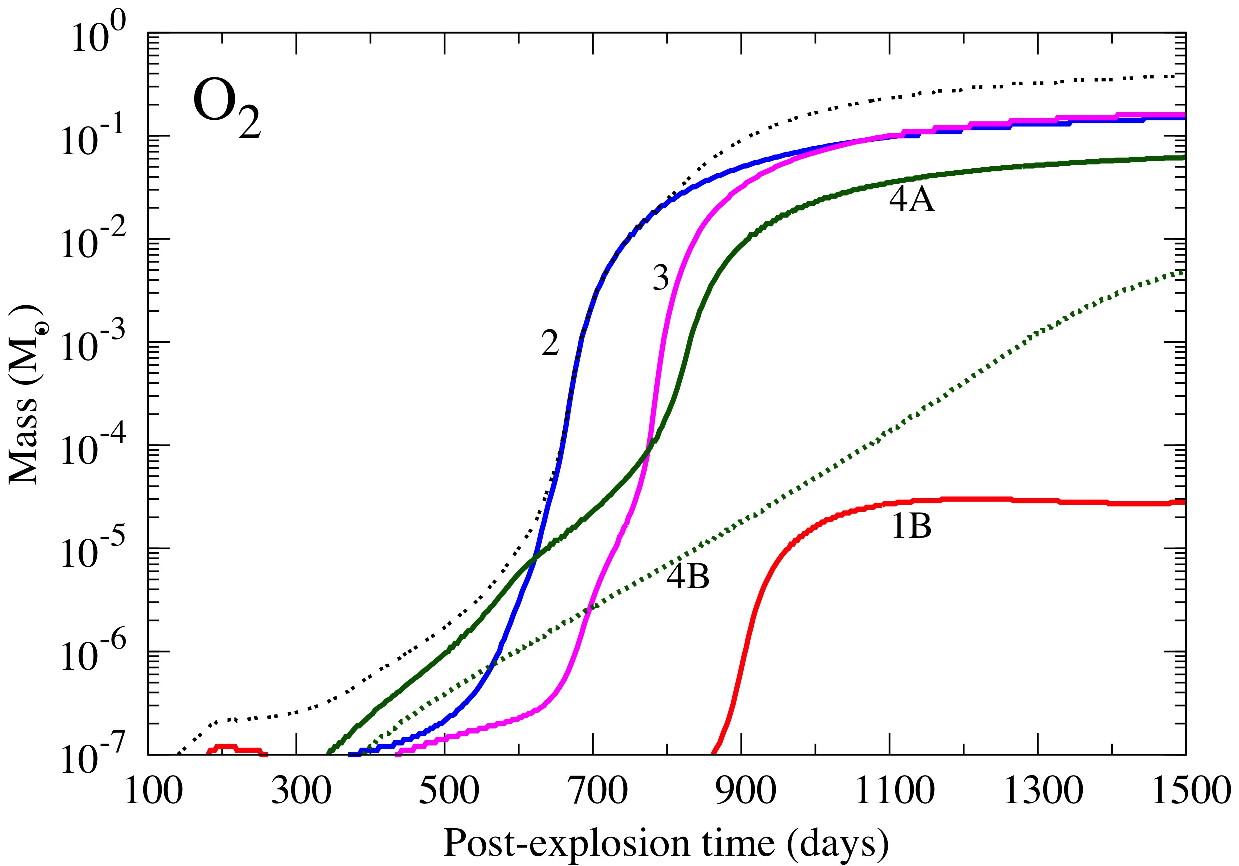}{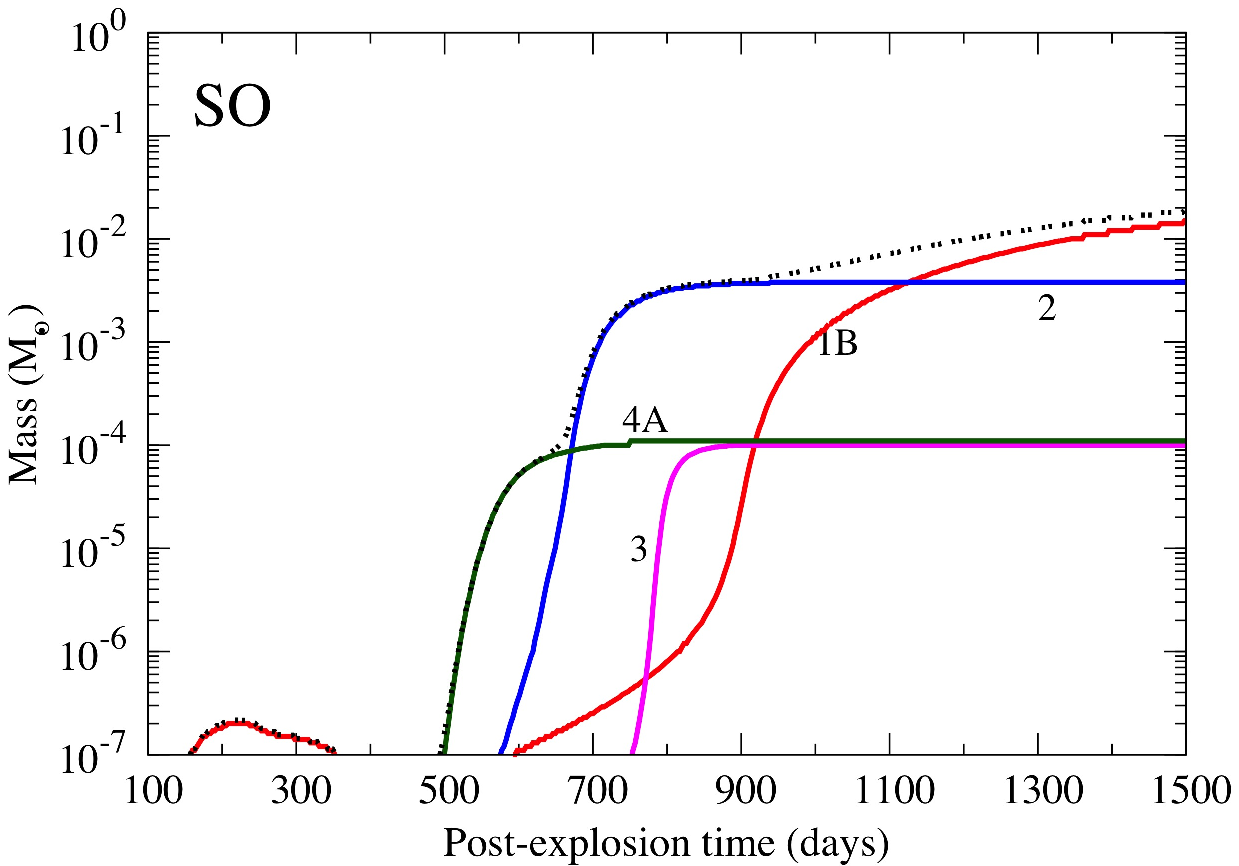}
\caption{Evolution of masses of O-bearing species with post-explosion time for the 15 \Ms~progenitor as a function of ejecta zones (see Table 1 for zone labelling): top) Mass of O$_2$; bottom) Mass of SO. The dotted-grey line represents the mass summed over all zones. \label{fig5}}
\end{figure}

\begin{figure}
\figurenum{6}
\epsscale{1.}
\plotone{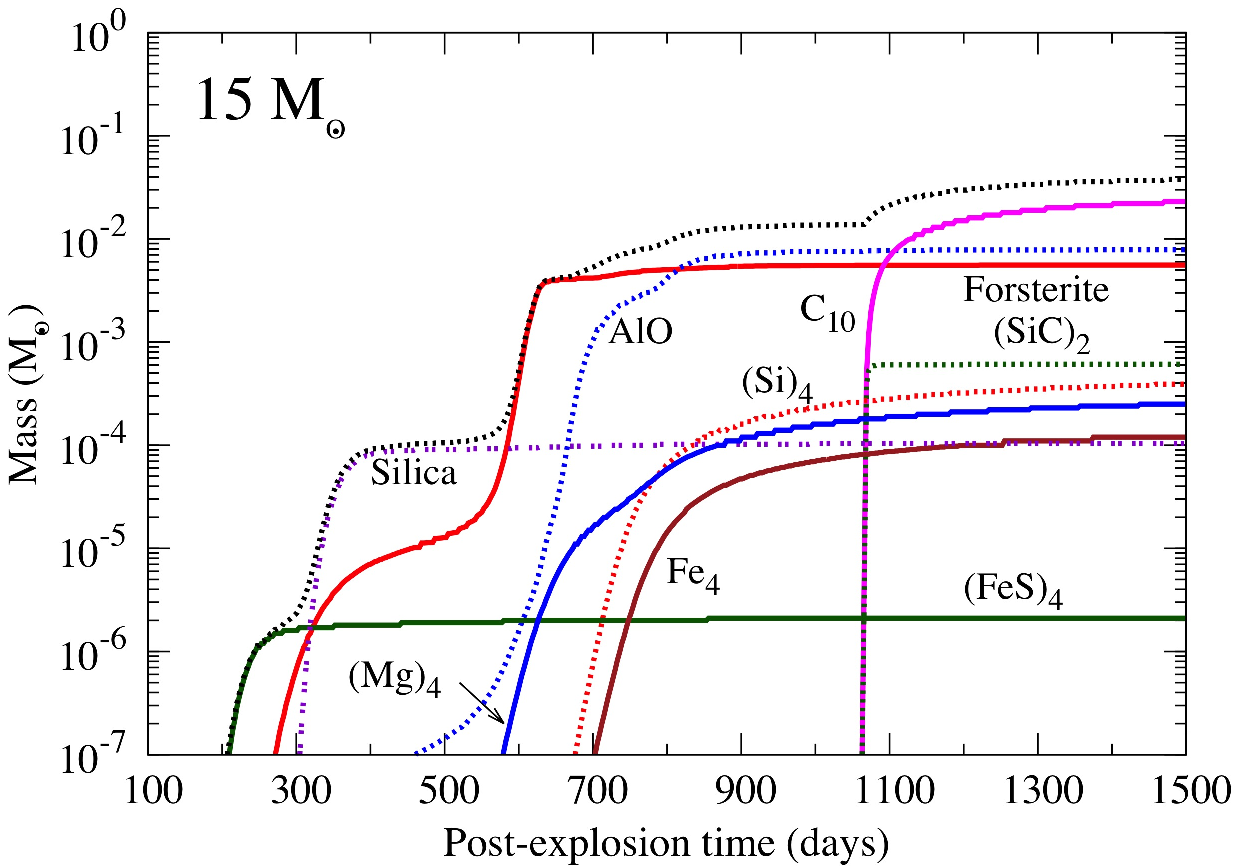}
\caption{Evolution of dust cluster masses with post-explosion time for the 15 \Ms~progenitor. For each cluster type, the masses have been summed over all ejecta zones. The dotted-grey line represents the total cluster mass, and provides an upper limit on the mass of dust that forms in the ejecta.\label{fig6}}
\end{figure}

\begin{figure}
\figurenum{7}
\epsscale{1.8}
\plottwo{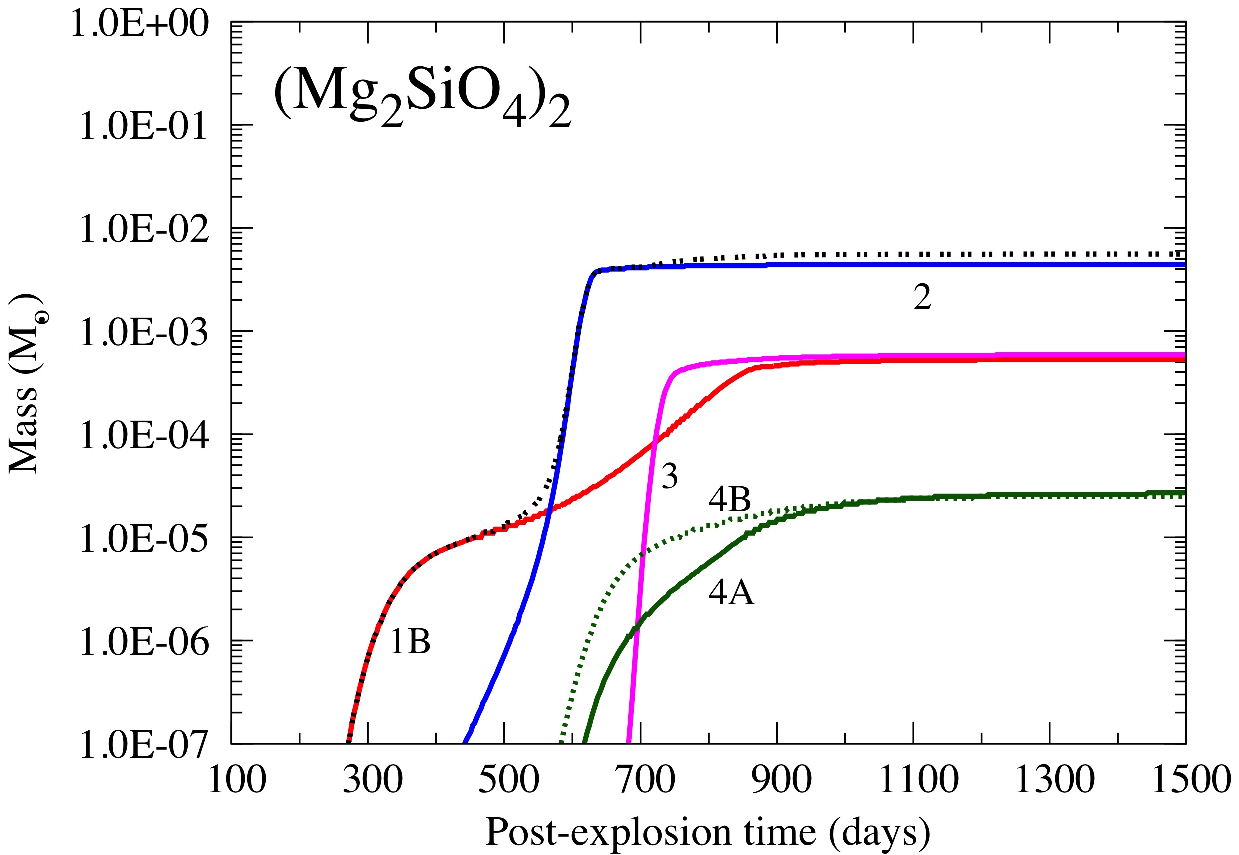}{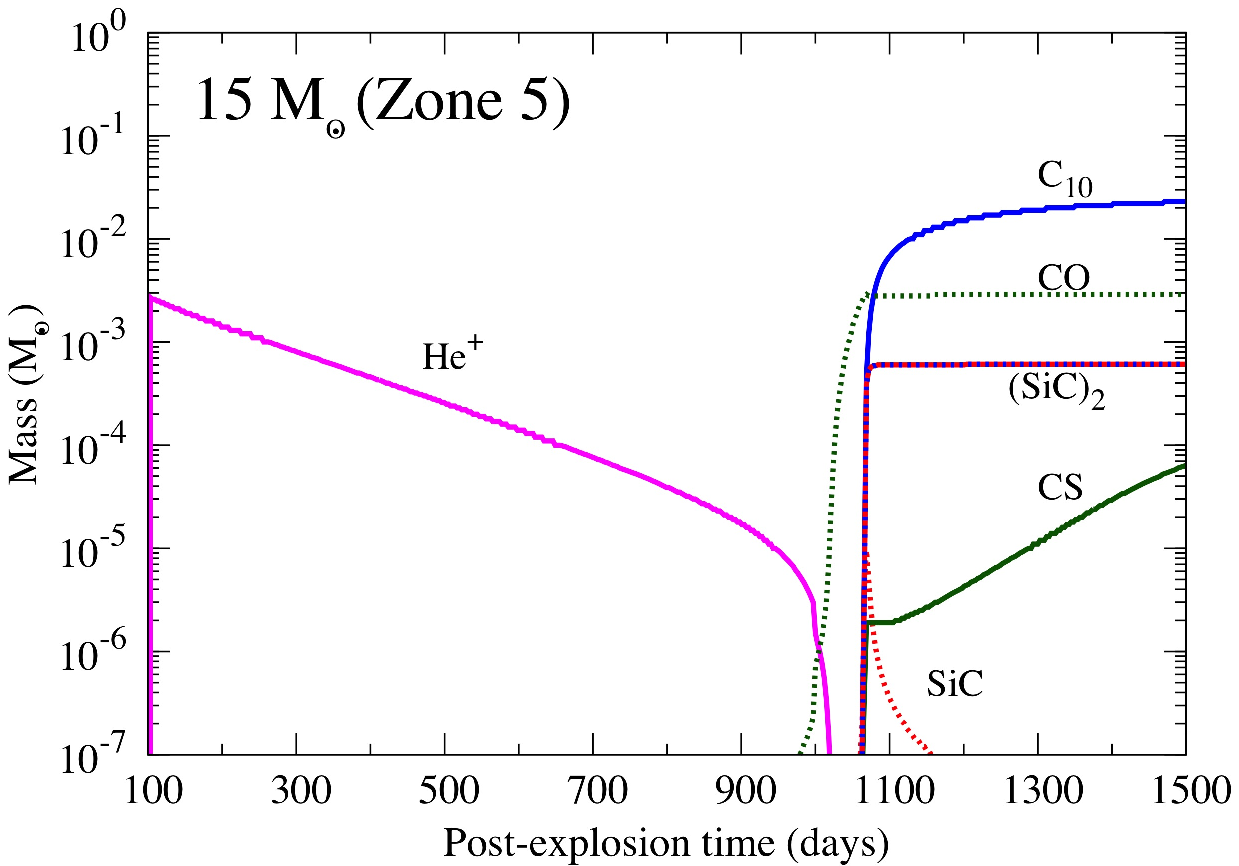}
\caption{Top: Forsterite dimer masses as a function of post-explosion time for the 15 \Ms~progenitor and the various ejecta zones (see Table 1 for zone labelling). Bottom: The carbon-rich cluster masses with post-explosion time in zone 5 of the 15 \Ms~progenitor. Clusters form once the \hep~masses decrease to negligible values after day 1000.\label{fig7} }
\end{figure}

\begin{figure}
\figurenum{8}
\epsscale{1.8}
\plottwo{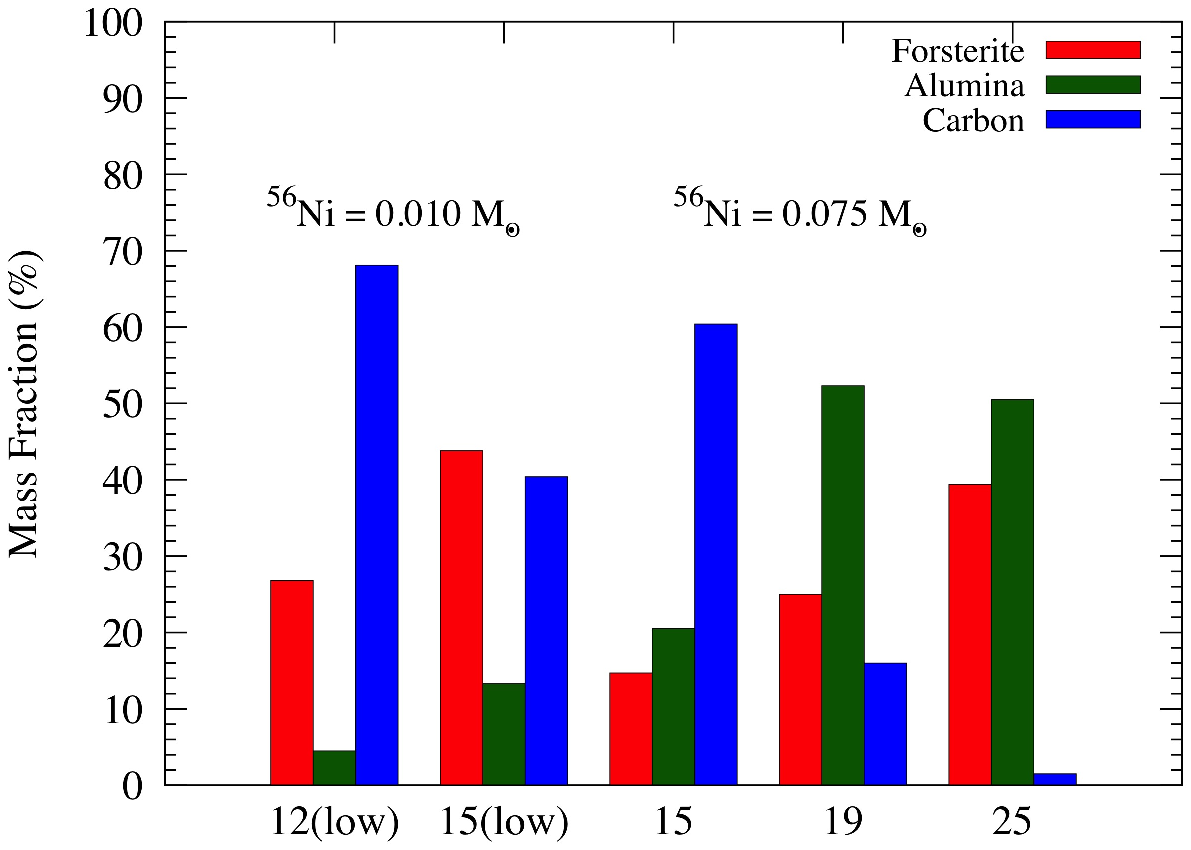}{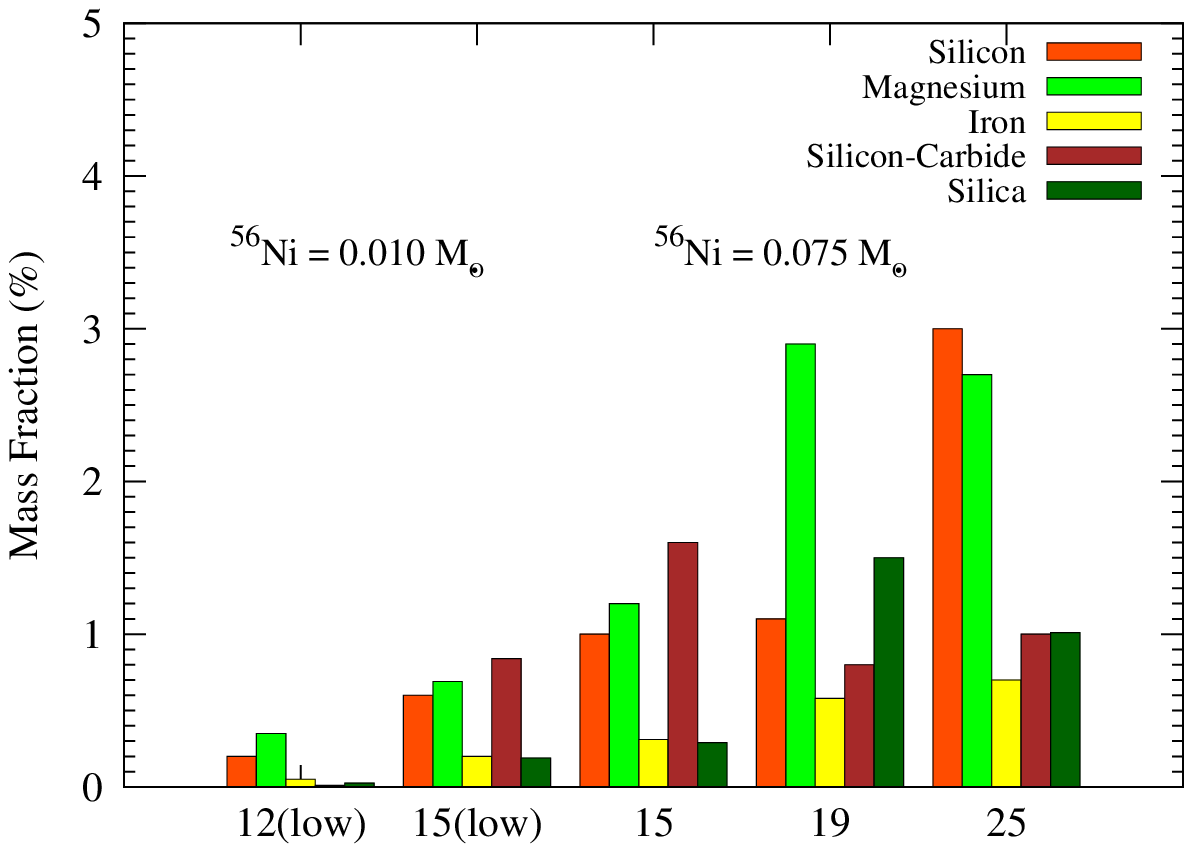}
\caption{Top: The mass fraction (in \%) of the major dust constituents entering the total dust mass produced at day 1500 versus progenitor mass. Bottom: The mass fraction (in \%) of the minor dust constituents entering the total dust mass produced versus progenitor mass (in \Ms). The total dust mass produced is 0.048 \Ms\ and 0.058 \Ms\ for, respectively, the 12 \Ms\ and 15 \Ms\ progenitors with low \Ni\ mass. For the 15, 19, and 25 \Ms\ progenitors with large \Ni\ mass, the total dust mass produced is 0.038, 0.035, and 0.09 \Ms, respectively. \label{fig8} }
\end{figure}

\begin{figure}
\figurenum{9}
\epsscale{1.8}
\plottwo{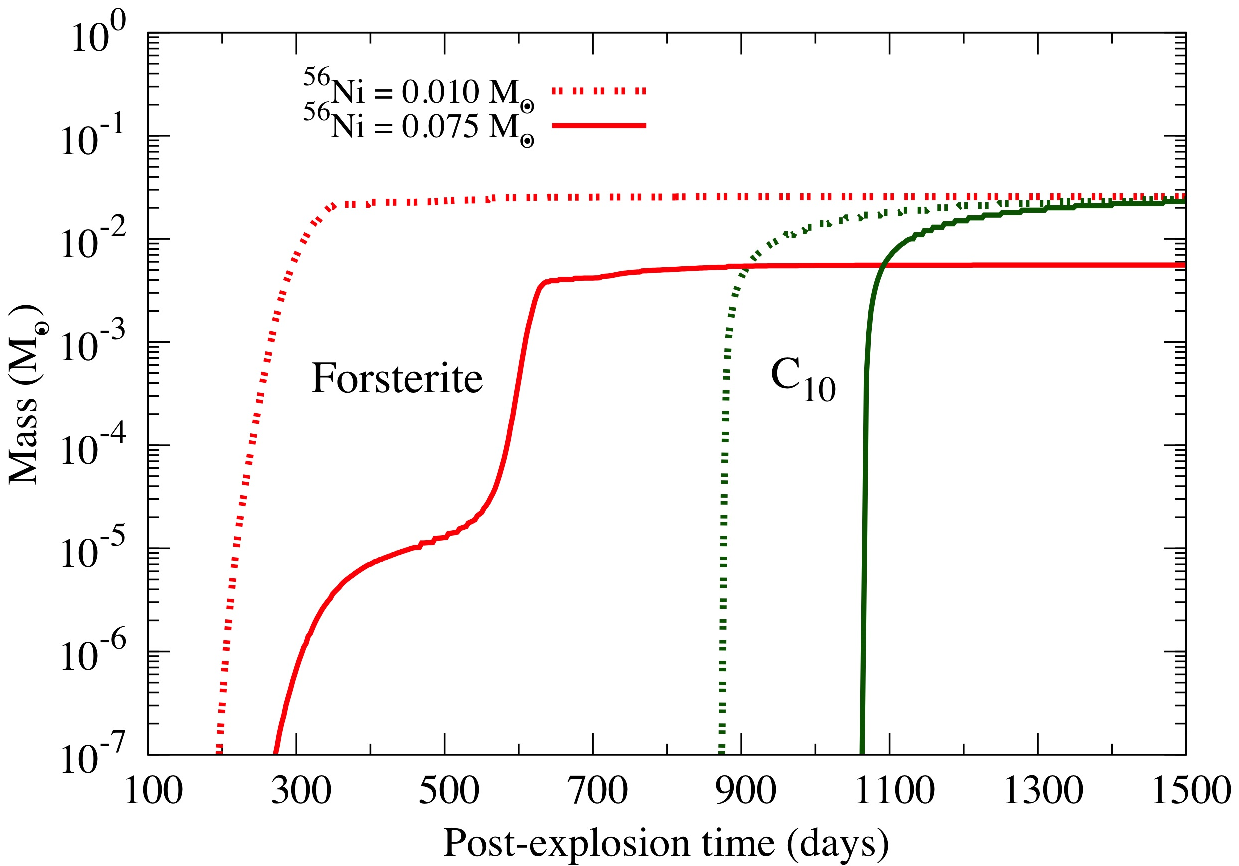}{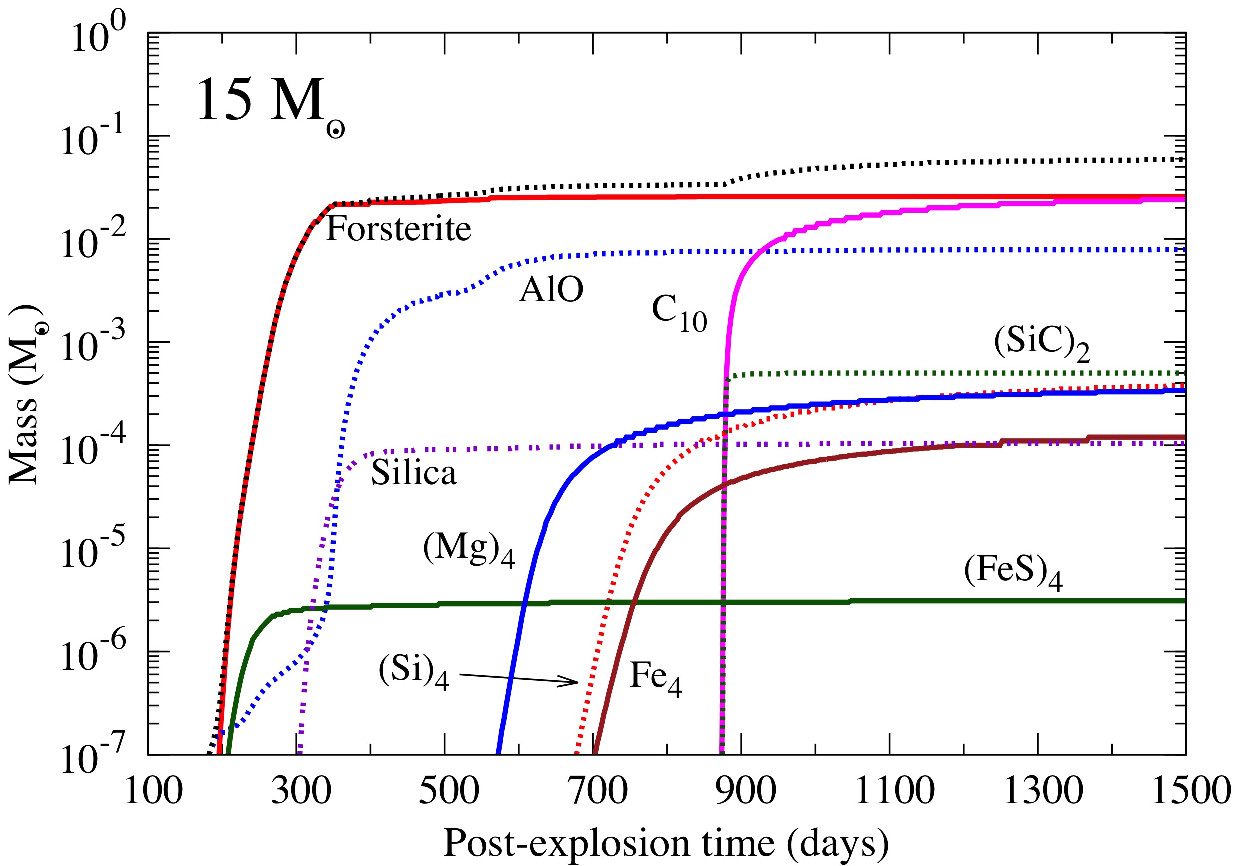}
\caption{Top: Mass of forsterite clusters and carbon rings for the 15 \Ms~progenitor as a function of post-explosion time and \Ni~mass. Bottom: Dust mass for the 15 \Ms~progenitor as a function of post-explosion time for a \Ni~mass of 0.01 \Ms. The dotted-grey line represents the total cluster mass. Dust clusters form at early time compared with the standard 15 \Ms~case (see Figure 6).\label{fig9} }
\end{figure}

\begin{figure}
\figurenum{10}
\epsscale{1.8}
\plottwo{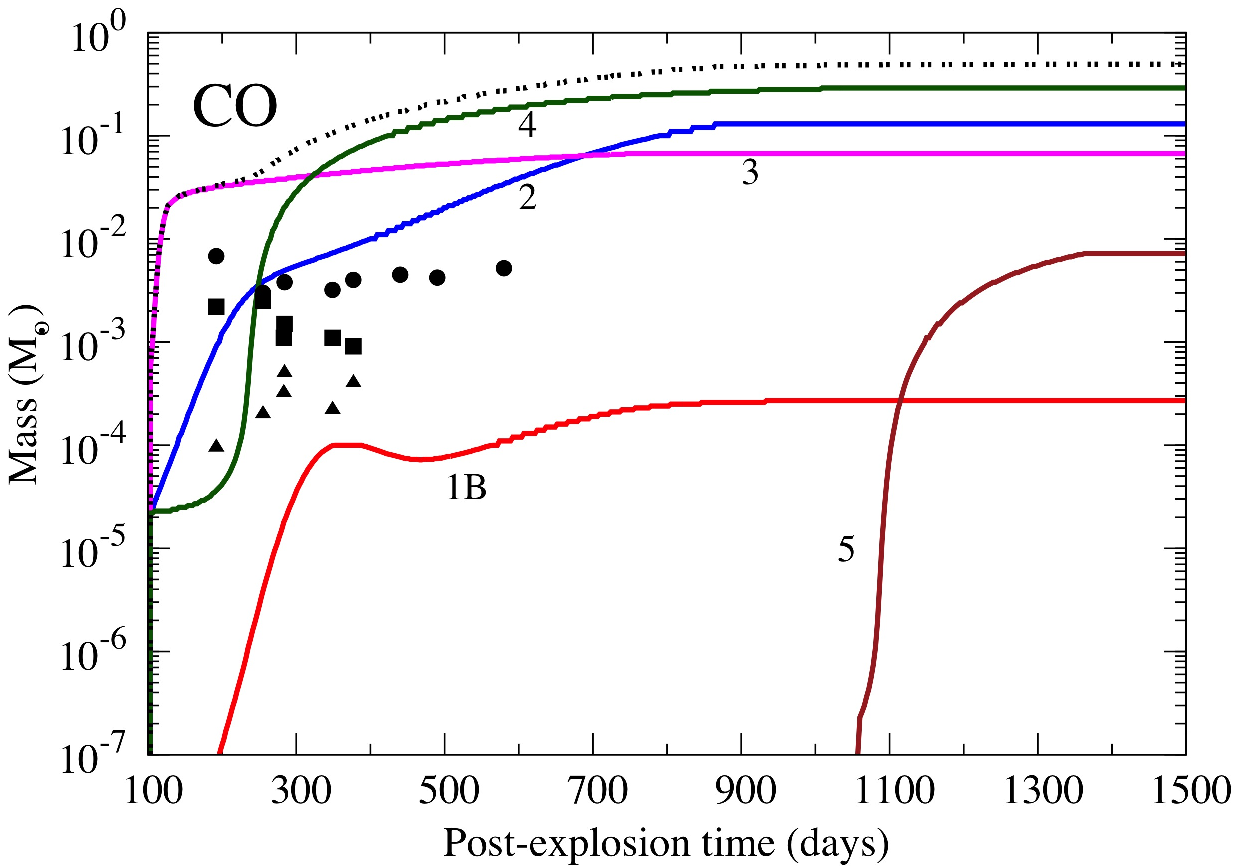}{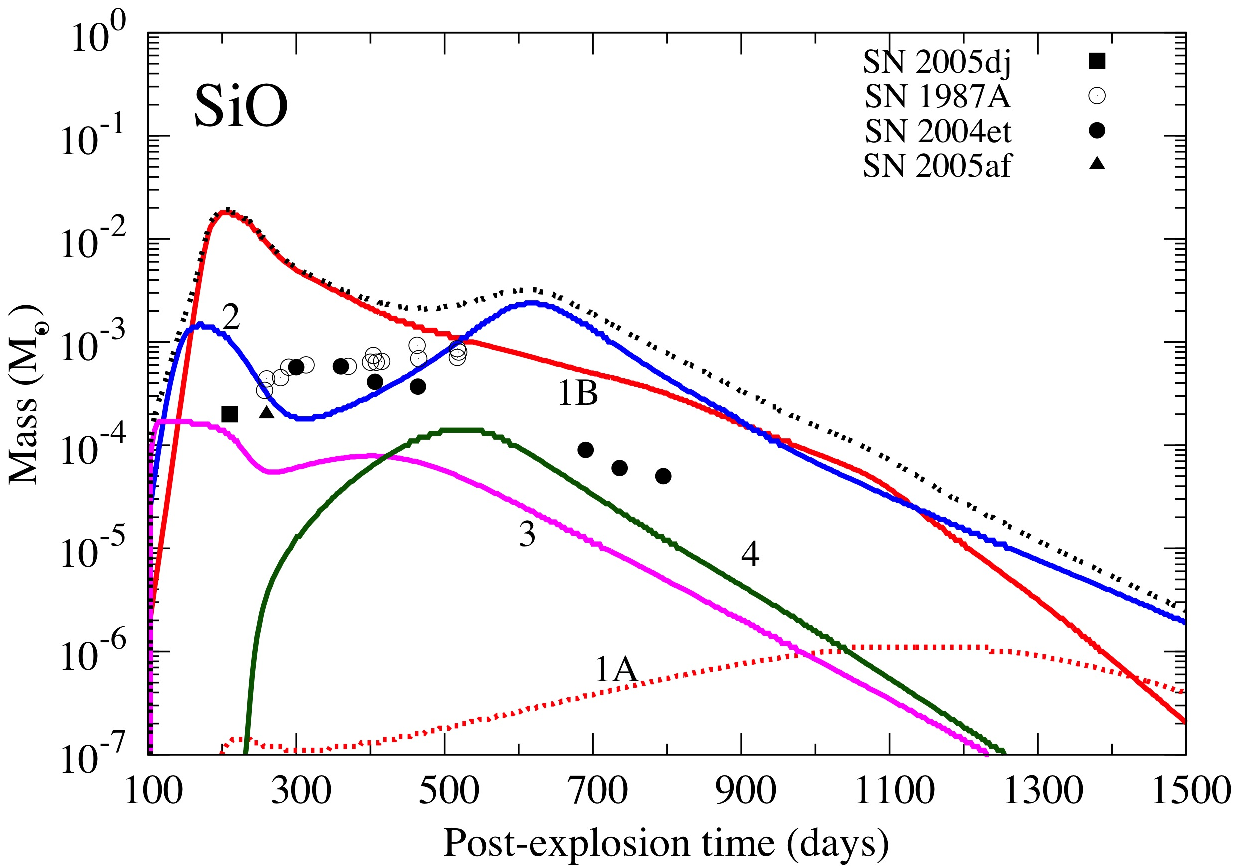}
\caption{Top: CO mass as a function of post-explosion time for the 19 \Ms~progenitor for the various ejecta zones (see Table \ref{tbl-1} for zone labelling). CO masses derived for SN1987A are also shown $-$ see Figure \ref{fig2} for details. Bottom: SiO mass evolution with post-explosion time for the 19 \Ms~progenitor. The masses derived for several SNe are also shown as symbols. The dotted-grey line represents the mass summed over all zones.\label{fig10} }
\end{figure}

\begin{figure}
\figurenum{11}
\epsscale{1.8}
\plottwo{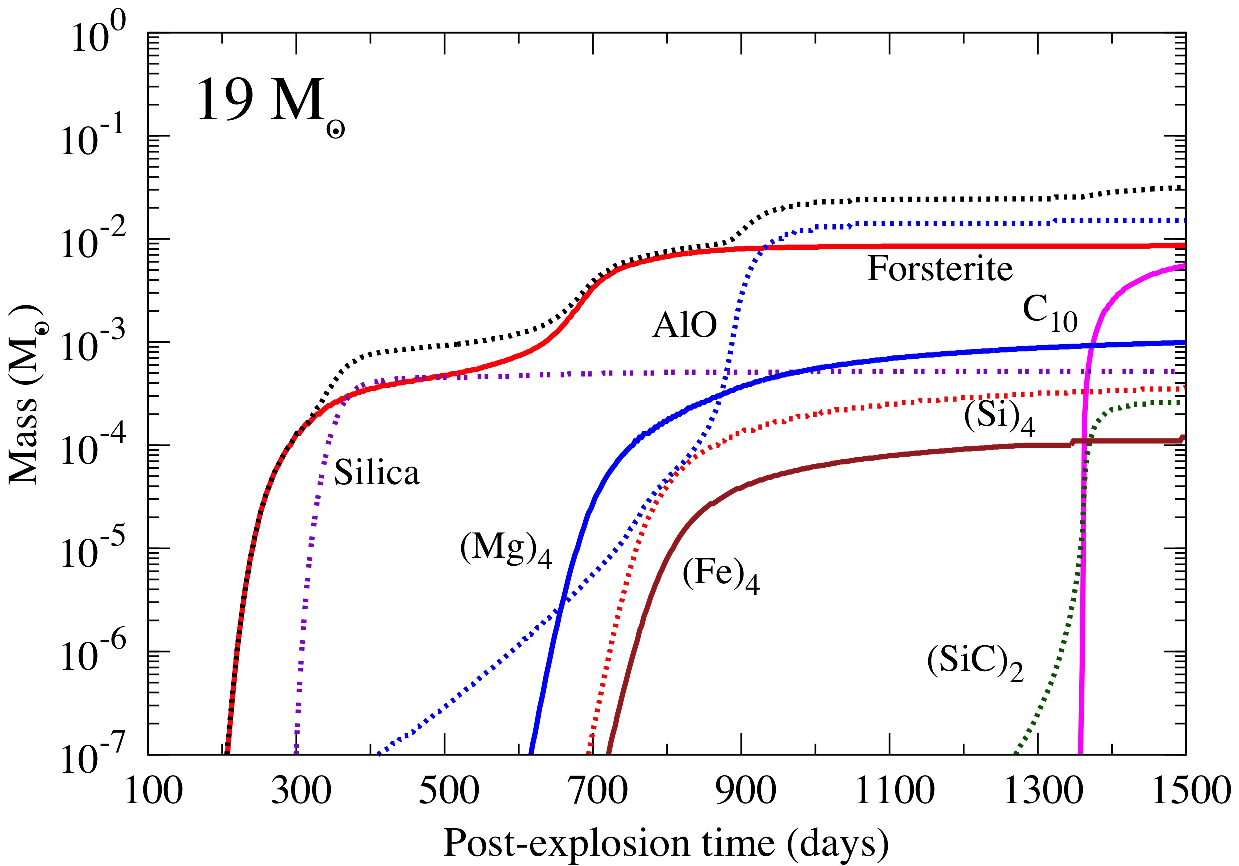}{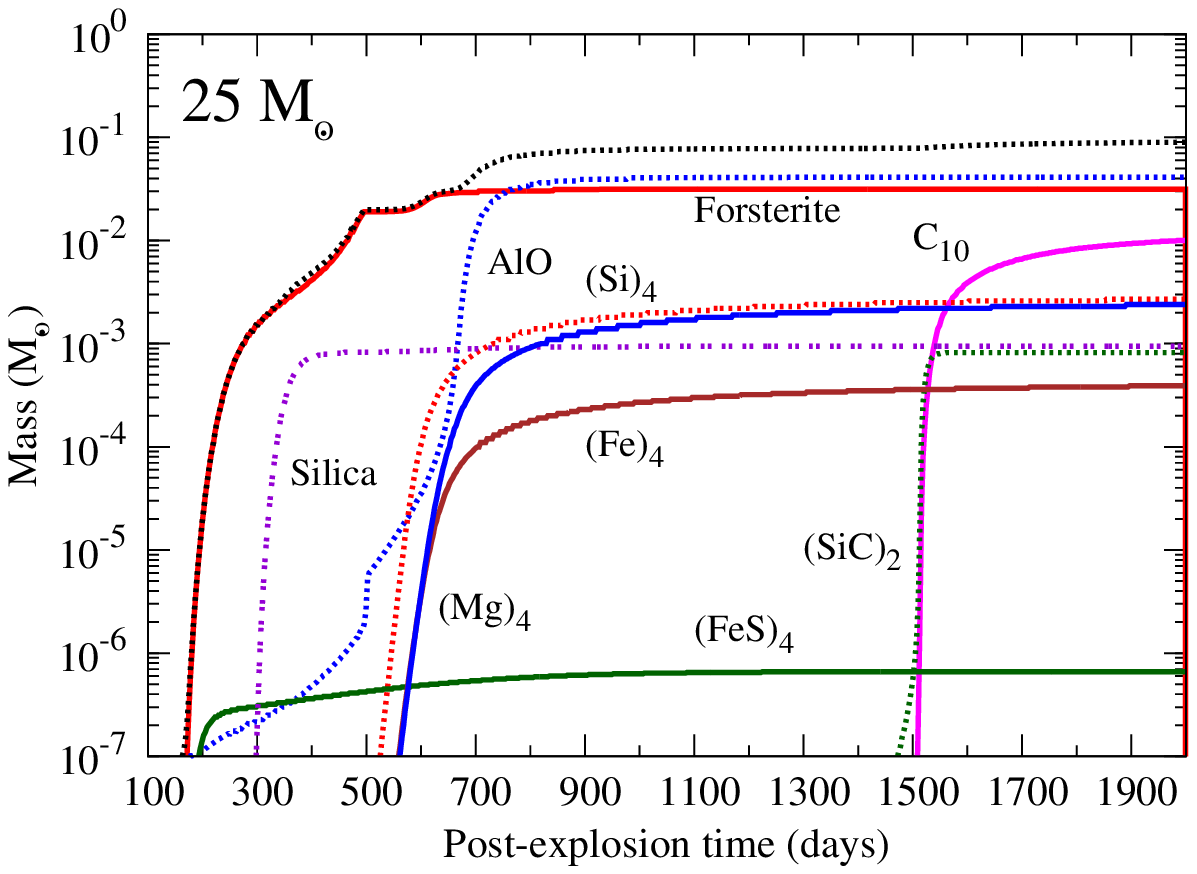}
\caption{Top: Evolution of dust cluster masses with post-explosion time for the 19 \Ms~progenitor. Bottom: Dust mass for the 25 \Ms~progenitor as a function of post-explosion time for a \Ni~mass of 0.075 \Ms. For each cluster type, the masses have been summed over all zones of the ejecta. The dotted-grey line represents the total cluster mass, and provides an upper limit on the mass of dust that forms in the ejecta. \label{fig11}}
\end{figure}

\begin{figure}
\figurenum{12}
\epsscale{1.}
\plotone{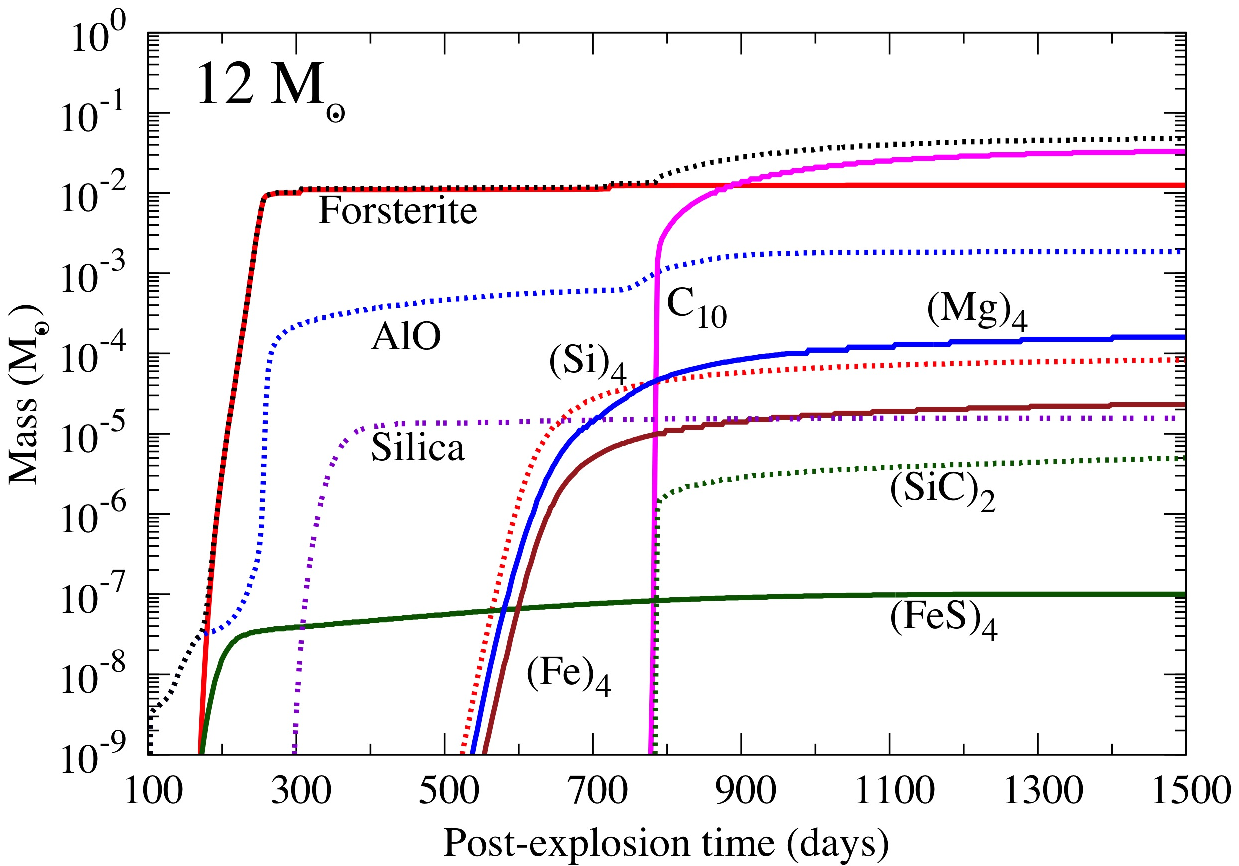}
\caption{Evolution of dust cluster masses with post-explosion time for the 12 \Ms~progenitor with a \Ni~mass of 0.01 \Ms. For each cluster type, the masses have been summed over all zones of the ejecta. The dotted-grey line represents the total cluster mass, and provides an upper limit on the mass of dust that forms in the ejecta. \label{fig12} }
\end{figure}

\begin{figure}
\figurenum{13}
\epsscale{1.8}
\plottwo{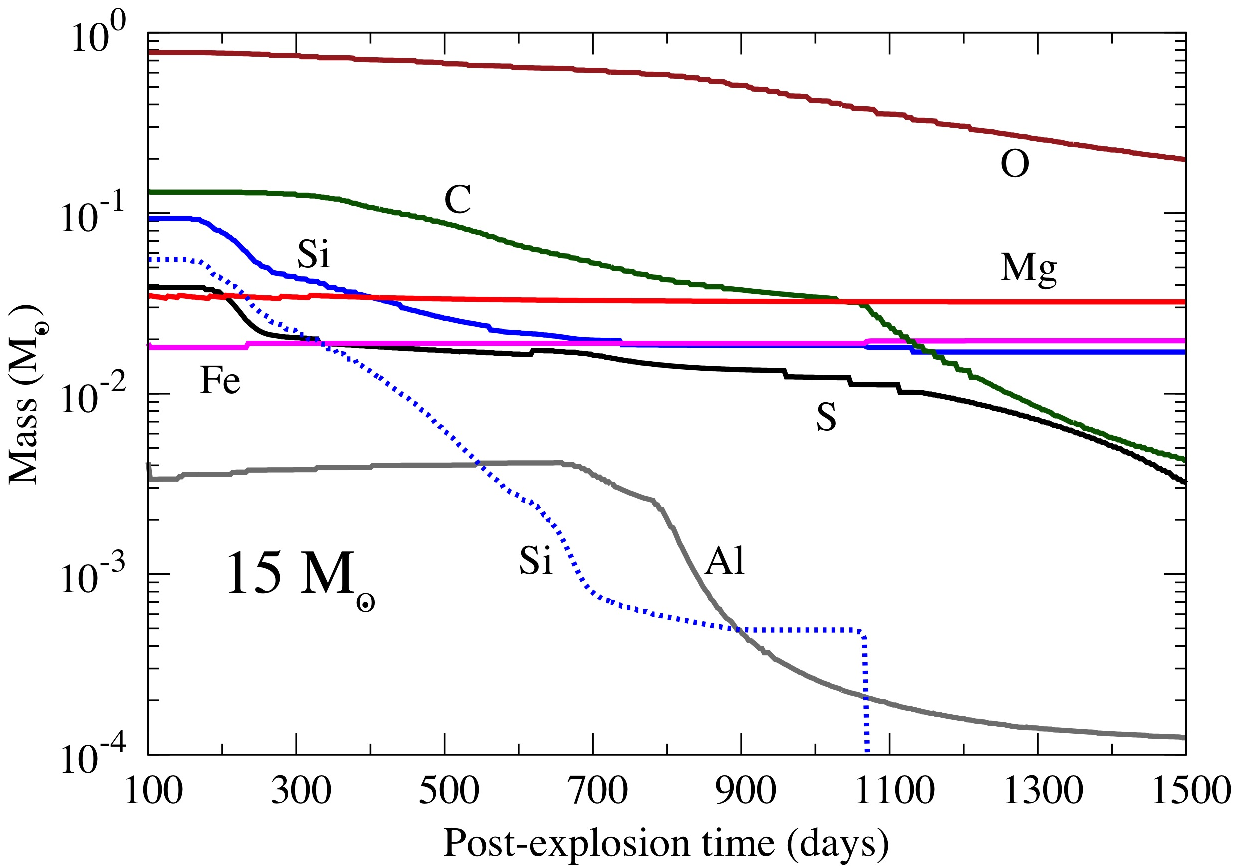}{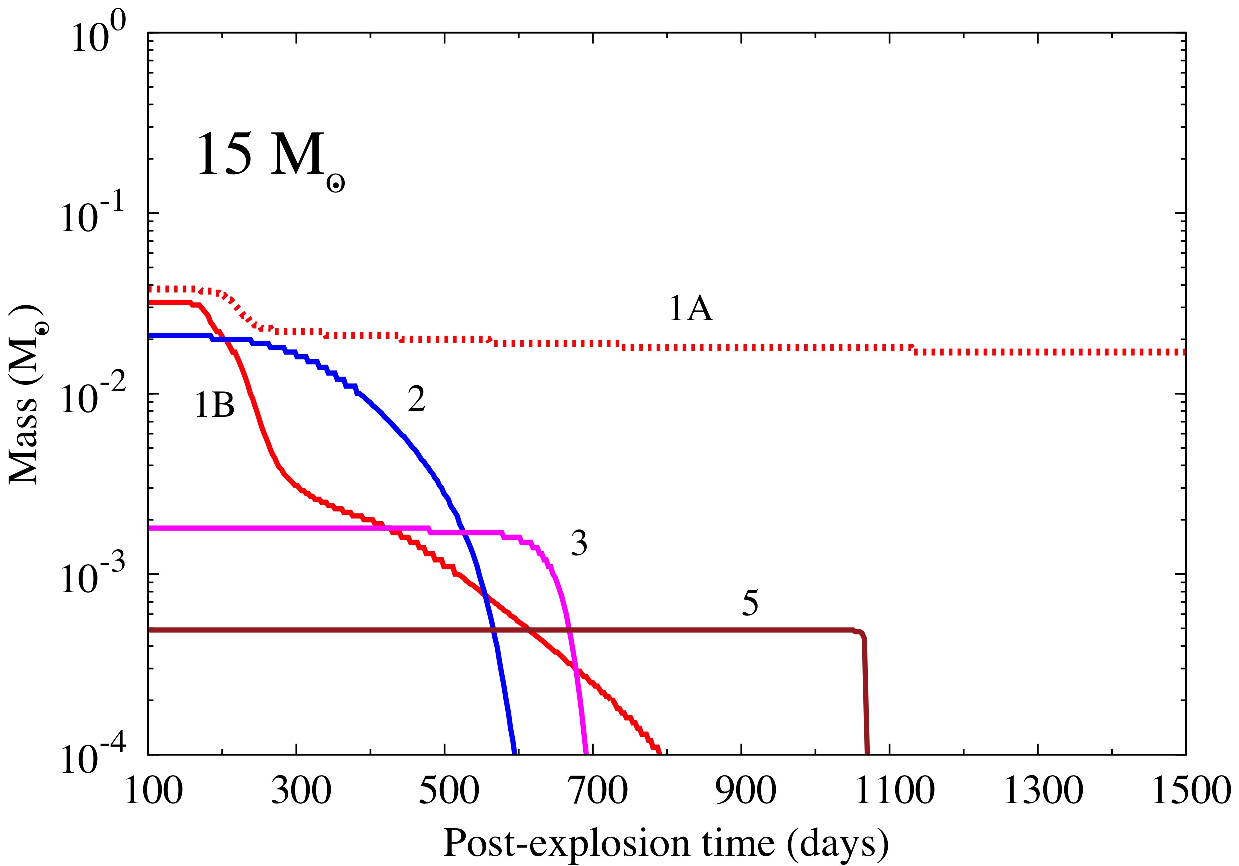}
\caption{Top: Evolution of element masses summed over all ejecta zones with post-explosion time for the 15 \Ms~progenitor. The dotted blue line represents the Si mass resulting from all zones except for the innermost zone, zone 1A. Bottom: Atomic silicon mass for the 15 \Ms~progenitor as a function of ejecta zones and post-explosion time. Si atoms are rapidly depleted in the formation of SiO, silica and silicate clusters in most of the zones except for zone 1A, where Si is primarily depleted in SiS. \label{fig13}}
\end{figure}

The present results on molecules may put constraint on the physical parameters of the ejecta. The formation of SiO dimers is a good example. The SiO dimer formation rate is gas pressure-dependent and usually very low at the low pressure encountered in the ejecta before day 400. When the SiO dimerisation rate derived by \citet{zac93} is used for the ejecta pressure, SiO and subsequent forsterite dimer formation is postponed to late epochs (t $>$ 700 days) as shown by Cherchneff \& Dwek (2010) for primeval, massive SN explosions. In the present models, the SiO dimerisation rate has been increased to account for the density enhancement found in clumps, and the match between SiO observational data and modelled masses is satisfactory (see Figure \ref{fig4}). We conclude that the observed SiO line fading and the timing for dust condensation are thus indirect indicators of the clumpy nature of SN ejecta.  

The upper limit on dust mass produced by our sample of SNe spans the $0.03 - 0.09 $ \Ms\ range. These values are much larger than those derived from IR data but somewhat less than values derived from submm data, i.e., 0.4$-$0.7 \Ms\ for SN1987A and 0.24 \Ms\ for the Crab Nebula. However, these large masses have been derived assuming a simple dust composition (usually, either carbon or silicate with the addition of some iron), and restricted physical parameters (i.e., one or two dust temperatures). The chemistry of dust synthesis has not been considered and condensing efficiencies of 100 \% of all available elements are usually assumed \citep[e.g.,][]{mat11}. All of these factors tend to boost the mass of solids synthesised in SN ejecta, when lower dust masses should be expected due to the bottleneck effect of the nucleation phase and the large variety of condensates produced in the gas. To validate our modelled dust chemical compositions, masses, and formation sequences, a study on the modelling of IR and submm fluxes for the homogeneous and clumpy ejecta of several SNe and SNRs will be presented in a forthcoming paper. 

The possibility of new formation of grains and their growth at late epochs (t $> 5$ yr) in the ejecta must be addressed. First, the formation of new grains in the expending ejecta after \env\ 5 yr should be hampered by the shortage of the chemical agents responsible for the first nucleation step, i.e. SiO or C$_2$, which are depleted in the ejecta between 300 and 2000 days, depending on the progenitor mass. Second, the growth of existing dust grains via accretion of abundant atoms or molecules such as atomic C, Mg, Si, or O$_2$ on the grain surface will happen on a time scale given by $ \tau_{ac}= [n_d\times \sigma_d \times v \times S(T, T_d)]^{-1} $ where $n_d$, $\sigma_d$, and $T_d$ are the number density, the collision cross section, and the temperature of the grains, respectively, $v$ and $T$ are the  thermal velocity and temperature of the gas, respectively, and $S(T, T_d)$ is the sticking coefficient. For typical grain sizes (0.1 \mic) and ejecta gas conditions after day 2000 ($n = 10^6$ cm$^{-3}$ and $T = 400$ K), the sticking coefficient is \env\ 0.5 and the estimated accretion time $\tau_{ac}$ is \env\ $10^4$ yr. This time scale exceeds the free expansion phase of SNe, and by that time, the ejecta will have been reprocessed by the reverse shock in the remnant. Therefore, late grain growth cannot proceed due to the very long accretion time required to add mass to the grains. The dust observed in SNe and SN remnants has thus formed in the nebular phase of the ejecta before \env\ 5 years after the explosion.

Finally, our finding of a gradual increase in dust mass due to a sequence of various condensation events in the ejecta reconciles the mass values derived from IR data with those from submm data. A hint of some increase in the dust mass over time was already indicated for SN1987A by \citet{wood93}, who inferred a 1.6 factor increase in the dust mass between 615 and 775 days, a value that agrees well with our results for the 19 \Ms\ progenitor. Obviously, the present models use simplistic, one-dimensional explosion models for SNe and have not yet included the dust condensation phase. In this regard, three-dimensional explosion models provide more realistic samples of clump chemical compositions. Because of the strong impact of \Ni\ on the ejecta chemistry through the formation of noble gas ions, each clump has a specific composition, thermal and density history, and thus a specific dust condensation scenario and efficiency. Applying a chemical kinetic formalism of the dust synthesis to such clumps, including the thermal feedback of molecules such as CO, SiO, and SiS, will fine-tune the prediction of the final dust mass produced by SN ejecta. However, from the present study, we anticipate that Type II-P SNe are efficient but moderate dust producers in local and remote galaxies. In the context of primeval galaxies at high$-r$edshift, the requirement that SNe produce \env\ 1 \Ms\ of dust in order to explain the large amount of dust produced at high redshift \citep{dwek07} is not satisfied. According to the present study, the explosion of primitive supergiant stars as SNe should produce at most 0.1 \Ms\ of O-rich dust, but grain destruction induced by shocks in the remnant phase will lower this value. These results argue for alternative and efficient O-rich and carbon dust providers (e.g., asymptotic giant branch stars, quasars) to account for the large dust masses present in the early universe.      

\acknowledgments We thank the anonymous referee for comments that have contributed to the improvement of the manuscript. We also thank Stefan Bromley, John Plane, Eli Dwek, Rubina Kotak, and Patrice Bouchet for fruitful discussions. A.S. acknowledges support from the Swiss National Science Foundation grant PMPD2-114347 attached to the ESF Eurogenesis network CoDustMas. 

\clearpage

\appendix
APPENDIX: \\
Two tables are provided in the Appendix. Table \ref{tbl-1} lists the chemical scheme and the reaction rates for the nucleation of clusters implicated in the nucleation phase. Table \ref{tbl-2} summarises the rates of reactions with the Compton electrons induced by radioactivity in the ejecta. 
\begin{deluxetable}{llclcccc}
\tabletypesize{\scriptsize}
\tablenum{A1}
\tablecaption{The chemical routes to nucleation of the various clusters considered in the present study. A pure chemical kinetic approach has been used whereby the reaction rates are either known (calculated from theory or measured in the laboratory) or estimated. Backward rates are not estimated from detailed balance and thermodynamic data (for more detail, see \citet{cher11a}. The reaction rates are expressed in Arrhenius form and the parameters for each reaction are indicated.\tablenotemark{a} \label{tbl-A1} }
 \tablewidth{0pt}
\tablehead{
Reaction& \multicolumn{2}{l}{Reactants} & Products& A$_{ij}$ & $\nu$ & E$_{a}$ & References\tablenotemark{b}  \\}
\startdata
 & \multicolumn{7}{c}{(SiO)$_n$ clusters} \\
\hline

A1 & SiO+SiO & $ \rightarrow $ & Si$_2$O$_2$ & 4.6086$ \times 10^{-17}$ & 0 & -2821.4 & ZT93, CD10\\
A2 & Si$_2$O$_2$+SiO & $ \rightarrow $ & Si$_3$O$_3$ & 2.2388$ \times 10^{-15}$ & 0 & -2878.9 & " \\
A3 & Si$_2$O$_2$+ Si$_2$O$_2$ & $ \rightarrow $ & Si$_3$O$_3$+SiO & 1.5265$ \times 10^{-14}$ & 0 & -2386.8 & " \\
A4 & Si$_3$O$_3$+SiO & $ \rightarrow $ & Si$_4$O$_4$ & 1.5265$ \times 10^{-14}$ & 0 & -2386.8 & " \\
A5 & Si$_2$O$_2$+ Si$_2$O$_2$ & $ \rightarrow $ & Si$_4$O$_4$ & 1.5265$ \times 10^{-14}$ & 0 & -2386.8 & " \\
A6 & Si$_3$O$_3$+ Si$_2$O$_2$ & $ \rightarrow $ & Si$_4$O$_4$+SiO & 1.5265$ \times 10^{-14}$ & 0 & -2386.8 & " \\
A7 & Si$_4$O$_4$+SiO & $ \rightarrow $ & Si$_5$O$_5$ & 1.5265$ \times 10^{-14}$ & 0 & -2386.8 & " \\
A8 & Si$_3$O$_3$+ Si$_2$O$_2$ & $ \rightarrow $ & Si$_5$O$_5$& 1.5265$ \times 10^{-14}$ & 0 & -2386.8 & " \\
A9 & Si$_2$O$_2$ & $ \rightarrow $ & SiO+SiO & 7.7200$ \times 10^{-7}$ & 0 & 0 & "\\ 
A10 & Si$_3$O$_3$ & $ \rightarrow $ & Si$_2$O$_2$+SiO & 7.8300$ \times 10^{-6}$ & 0 & 0 & "\\
A11 & Si$_4$O$_4$ & $ \rightarrow $ & Si$_3$O$_3$+SiO & 9.9000$ \times 10^{-4}$ & 0 & 0 & "\\
A12 & Si$_4$O$_4$ & $ \rightarrow $ & Si$_2$O$_2$+Si$_2$O$_2$ & 9.9000$ \times 10^{-4}$ & 0 & 0 & "\\
A13 & Si$_5$O$_5$ & $ \rightarrow $ & Si$_3$O$_3$+Si$_2$O$_2$ & 9.9000$ \times 10^{-4}$ & 0 & 0 & "\\

\hline
 & \multicolumn{7}{c}{Forsterite (Mg$_4$Si$_2$O$_8$) and Enstatite (Mg$_2$Si$_2$O$_6$) dimers} \\
\hline
B1 & Si$_2$O$_2$+O$_2$ & $\rightarrow$ & Si$_2$O$_3$+O & 1.0000$ \times 10^{-11}$ & 0 & 1000 & E\\
B2 & Si$_2$O$_2$+SO & $\rightarrow$ & Si$_2$O$_3$+S & 1.0000$ \times 10^{-11}$ & 0 & 1000 & as B1 \\
B3 & Si$_2$O$_3$+Mg & $\rightarrow$ & MgSi$_2$O$_3$ &  1.0000$ \times 10^{-12}$ & 0 & 0 & E \\
B4 & MgSi$_2$O$_3$+O$_2$ & $\rightarrow$ & MgSi$_2$O$_4$+O &  1.0000$ \times 10^{-12}$ & 0 & 0  & as B3\\
B5 & MgSi$_2$O$_3$+SO & $\rightarrow$ & MgSi$_2$O$_4$+S &  1.0000$ \times 10^{-12}$ & 0 & 0  & as B3\\
B6 & MgSi$_2$O$_4$+Mg & $\rightarrow$ & Mg$_2$Si$_2$O$_4$ &  1.0000$ \times 10^{-12}$ & 0 & 0  & as B3 \\
B7 & Mg$_2$Si$_2$O$_4$+O$_2$ & $\rightarrow$ & Mg$_2$Si$_2$O$_5$+O & 1.0000$ \times 10^{-12}$ & 0 & 0  & as B3 \\
B8 & Mg$_2$Si$_2$O$_4$+SO & $\rightarrow$ & Mg$_2$Si$_2$O$_5$+S &  1.0000$ \times 10^{-12}$ & 0 & 0  & as B3\\
B9 & Mg$_2$Si$_2$O$_5$+O$_2$ & $\rightarrow$ & Mg$_2$Si$_2$O$_6$+O &  1.0000$ \times 10^{-12}$ & 0 & 0  & as B3 \\
B10 & Mg$_2$Si$_2$O$_5$+SO & $\rightarrow$ & Mg$_2$Si$_2$O$_6$+S &  1.0000$ \times 10^{-12}$ & 0 & 0  & as B3 \\
B11 & Mg$_2$Si$_2$O$_6$+Mg & $\rightarrow$ & Mg$_3$Si$_2$O$_6$ &  1.0000$ \times 10^{-12}$ & 0 & 0  & as B3 \\
B12 & Mg$_3$Si$_2$O$_6$+O$_2$ & $\rightarrow$ & Mg$_3$Si$_2$O$_7$+O & 1.0000$ \times 10^{-12}$ & 0 & 0 & as B3 \\
B13 & Mg$_3$Si$_2$O$_6$+SO & $\rightarrow$ & Mg$_3$Si$_2$O$_7$+S &  1.0000$ \times 10^{-12}$ & 0 & 0  & as B3 \\
\\\\\\
B14 & Mg$_3$Si$_2$O$_7$+Mg & $\rightarrow$ & Mg$_4$Si$_2$O$_7$ & 1.0000$ \times 10^{-12}$ & 0 & 0  & as B3 \\
B15 & Mg$_4$Si$_2$O$_7$+O$_2$ & $\rightarrow$ & Mg$_4$Si$_2$O$_8$+O &  1.0000$ \times 10^{-12}$ & 0 & 0  & as B3 \\
B16 & Mg$_4$Si$_2$O$_7$+SO & $\rightarrow$ & Mg$_4$Si$_2$O$_8$+S &  1.0000$ \times 10^{-12}$ & 0 & 0  & as B3 \\
\hline 
 & \multicolumn{7}{c}{Si$_n$O$_{n+1}$ clusters} \\
\hline
C1 & Si$_2$O$_3$+O & $\rightarrow$ & Si$_2$O$_2$+O$_2$ & 1.0000$ \times 10^{-12}$ & 0 & 0  & E \\
C4 & Si$_2$O$_3$+S & $\rightarrow$ & Si$_2$O$_2$+SO &  1.0000$ \times 10^{-12}$ & 0 & 0  & E \\
C5 & Si$_3$O$_3$+O$_2$ & $\rightarrow$ & Si$_3$O$_4$+O &  1.0000$ \times 10^{-13}$ & 0 & 1000   & as B1 \\
C6 & Si$_3$O$_3$+SO & $\rightarrow$ & Si$_3$O$_4$+S & 1.0000$ \times 10^{-13}$ & 0 & 1000   & '' \\
C7 & Si$_4$O$_4$+O$_2$ & $\rightarrow$ & Si$_4$O$_5$+O &  1.0000$ \times 10^{-13}$ & 0 & 1000   & " \\
C8 & Si$_4$O$_4$+SO & $\rightarrow$ & Si$_4$O$_5$+S &  1.0000$ \times 10^{-13}$ & 0 & 1000   & '' \\
C9 & Si$_2$O$_3$+SiO & $ \rightarrow $ & Si$_3$O$_4$ & 7.4627$ \times 10^{-16}$ & 0 & -2878.9 & ZT93 \\
C10 & Si$_3$O$_4$+SiO & $ \rightarrow $ & Si$_4$O$_5$ & 5.0884$ \times 10^{-15}$ & 0 & -2386.8 & " \\
C11 & Si$_2$O$_2$+SiO & $\rightarrow$ & Si$_2$O$_3$+Si &  7.4627$ \times 10^{-16}$ & 0 & -2878.9   & "\\
C12 & Si$_3$O$_3$+SiO & $\rightarrow$ & Si$_3$O$_4$+Si &  5.0884$ \times 10^{-15}$ & 0 & -2386.8 & "\\
C13 & Si$_4$O$_4$+SiO & $\rightarrow$ & Si$_4$O$_5$+Si &  5.0884$ \times 10^{-15}$ & 0 & -2386.8  & "\\
\hline 
& \multicolumn{7}{c}{Cluster Fragmentation} \\
\hline
D1 & Si$_2$O$_2$+M & $\rightarrow$ & SiO+SiO+M & 4.4000$ \times 10^{-10}$ & 0 & 98600.0 & CD10  \\
D2& Si$_3$O$_3$+M & $\rightarrow$ &Si$_2$O$_2$+SiO+M &4.4000$ \times 10^{-10}$&0& 98600.0 &  " \\
D3& Si$_4$O$_4$+M & $\rightarrow$ &Si$_3$O$_3$+SiO+M &4.4000$ \times 10^{-10}$&0& 98600.0  & " \\
D4 & Si$_4$O$_4$+M & $\rightarrow$ &Si$_2$O$_2$+Si$_2$O$_2$+M &4.4000$ \times 10^{-10}$&0& 98600.0 & " \\
D5 & Si$_5$O$_5$+M & $\rightarrow$ &Si$_4$O$_4$+SiO+M &4.4000$ \times 10^{-10}$&0& 98600.0 & " \\
D6 & Si$_5$O$_5$+M & $\rightarrow$ &Si$_2$O$_2$+Si$_3$O$_3$+M &4.4000$ \times 10^{-10}$&0& 98600.0 & " \\
D7& Si$_2$O$_3$+M & $\rightarrow$ & Si$_2$O$_2$+O+M &5.0000$ \times 10^{-10}$ & 0& 55000.0 & " \\
D8& Si$_3$O$_4$+M & $\rightarrow$ & Si$_3$O$_3$+O+M &5.0000$ \times 10^{-10}$ &0 & 55000.0  & " \\
D9& Si$_4$O$_5$+M & $\rightarrow$ & Si$_4$O$_4$+O+M &5.0000$ \times 10^{-10}$ &0& 55000.0 &  " \\
D10 & MgSi$_2$O$_3$+M & $\rightarrow$ & Si$_2$O$_3$+Mg+M & 1.0000$ \times 10^{-10}$ &0& 98600.0 &  as D1\\
D11 & MgSi$_2$O$_4$+M & $\rightarrow$ & MgSi$_2$O$_3$+O+M &1.0000$ \times 10^{-10}$ &0&98600.0 &  " \\
D12& Mg$_2$Si$_2$O$_4$+M & $\rightarrow$ & MgSi$_2$O$_4$+Mg+M&1.0000$ \times 10^{-10}$&0 & 98600.0 & "\\
D13 & Mg$_2$Si$_2$O$_5$+M & $\rightarrow$ & Mg$_2$Si$_2$O$_4$+O+M&1.0000$ \times 10^{-10}$&0 & 98600.0  &  " \\
D14 & Mg$_2$Si$_2$O$_6$+M & $\rightarrow$ & Mg$_2$Si$_2$O$_5$+O+M &1.0000$ \times 10^{-10}$&0& 98600.0 &  " \\
D15& Mg$_3$Si$_2$O$_6$+M & $\rightarrow$ & Mg$_2$Si$_2$O$_6$+Mg+M &1.0000$ \times 10^{-10}$&0& 98600.0 & " \\
D16 & Mg$_3$Si$_2$O$_7$+M & $\rightarrow$ & Mg$_3$Si$_2$O$_6$+O+M &1.0000$ \times 10^{-10}$&0& 98600.0 & " \\
D17& Mg$_4$Si$_2$O$_7$+M & $\rightarrow$ & Mg$_3$Si$_2$O$_7$+Mg+M &1.0000$ \times 10^{-10}$&0& 98600.0 & " \\
D18 & Mg$_4$Si$_2$O$_8$+M & $\rightarrow$ & Mg$_4$Si$_2$O$_7$+O+M &1.0000$ \times 10^{-10}$&0& 98600.0 &  " \\
\enddata
\tablenotetext{a}{Rates are given in the Arrhenius form $k= A_{ij} \times (T/300 K)^{\nu} \times \exp(-E_a/T)$ with $A_{ij}$ in s$^{-1}$, cm$^3$ s$^{-1}$, or cm$^6$ s$^{-1}$ for uni-, bi- and ter-molecular processes, and $E_a$ in Kelvin.}
\tablenotetext{b}{ZT93 $\equiv$ Zachariah \& Tsang (1993); CD10 $\equiv$ Cherchneff \& Dwek (2010); E $\equiv$ Estimated. }
\end{deluxetable}
\clearpage

\begin{deluxetable}{clccc|cccc}
\rotate
\tabletypesize{\scriptsize}
\tablenum{A2}
\tablecaption{Compton electron-induced reactions, corresponding mean energy per ion pair $W_{i}$ and Arrhenius coefficient A as a function of ejecta  model.\label{tbl-A2}                                                                                                                                                                                                                                                                                                                                                                                                                                                                                                                                                                                                                                                                                                                                                                                                                                                                                                                                                                                                                                                                                                                                                                                                                                                                      }

 \tablewidth{0pt}
\tablehead{
&&& \multicolumn{2}{c}{$^{56}Ni$ = 0.07 \Ms} & \multicolumn{3}{c}{$^{56}Ni$ = 0.01 \Ms} & \\
}

\startdata
Species & Reactions & W$_{i}$ (eV) & A - 15 \Ms \tablenotemark{a} & A - 19 \Ms\tablenotemark{a} & A - 12 \Ms\tablenotemark{a} & A - 15 \Ms\tablenotemark{a} & A - 25 \Ms\tablenotemark{a} & Reference\\
\tableline
CO &$ \rightarrow$ O$^+$ + C& 768 &7.7671$ \times 10^{-7} $ &6.1663$ \times 10^{-7} $ & 4.9768$\times 10^{-7}$&1.1763$ \times 10^{-7} $& 5.6102$ \times 10^{-8} $& Liu \& Dalgarno (1995)\\
 & $ \rightarrow$ C$^+$ + O &247 & 2.4150$ \times 10^{-6} $ &1.9173$ \times 10^{-6} $ &1.5472$ \times 10^{-6} $&3.6576$ \times 10^{-7} $&1.7444$ \times 10^{-7} $& "  \\
 & $ \rightarrow$ C + O & 125& 4.7722$ \times 10^{-6}$ &3.7887$ \times 10^{-6} $ & 3.0575$ \times 10^{-6}$&7.2268$ \times 10^{-7} $ & 3.4466$ \times 10^{-7} $&" \\
 & $ \rightarrow$ CO$^+$ + e$^-$ & 34 & 1.7544$ \times 10^{-5}$ &1.3928$ \times 10^{-5} $ & 1.1241$ \times 10^{-5}$&2.6570$ \times 10^{-6} $ & 1.2672$ \times 10^{-6} $&" \\
 \tableline
O & $ \rightarrow$ O$^+$ + e$^-$ & 46.2 & 1.2911$ \times 10^{-5}$ &1.3082$ \times 10^{-5} $ & 8.2723$ \times 10^{-6}$ &1.9554$ \times 10^{-6} $&9.3259$ \times 10^{-7} $& " \\
\tableline
C& $ \rightarrow$ C$^+$ + e$^-$ & 36.4 & 1.6297$ \times 10^{-5}$ &1.3010$ \times 10^{-5} $ & 1.0500$ \times 10^{-5}$&2.4819$ \times 10^{-6} $ &1.1837$ \times 10^{-6} $& " \\
\tableline
SiO& $ \rightarrow$ O$^+$ + Si & 678 & 8.7986$ \times 10^{-7}$ &6.9852$ \times 10^{-7} $ &5.6372$ \times 10^{-7}$&1.3324$ \times 10^{-7} $ &6.3546$ \times 10^{-8} $& " \\
& $ \rightarrow$ Si$^+$ + O &218 & 2.7363$ \times 10^{-6} $ &2.1724$ \times 10^{-6} $ &1.7531$ \times 10^{-6} $&4.1441$ \times 10^{-7} $&1.9764$ \times 10^{-7} $& "  \\
& $ \rightarrow$ Si + O &110 & 5.4228$ \times 10^{-6} $ &4.3051$ \times 10^{-6} $ &3.4747$ \times 10^{-6} $&8.2128$ \times 10^{-7} $& 3.9169$ \times 10^{-7} $&"  \\
& $ \rightarrow$ SiO$^+$ + e$^-$ &30 & 1.9884$ \times 10^{-5} $ &1.5786$ \times 10^{-5} $ &1.2740$ \times 10^{-5} $&3.0114$ \times 10^{-6} $&1.4362$ \times 10^{-6} $& "  \\
\tableline
N$_2$& $ \rightarrow$ N$^+$ + N & 264 & 2.2594$ \times 10^{-6}$ &1.7938$ \times 10^{-6} $ &1.4477$ \times 10^{-6}$ &3.4219$ \times 10^{-7} $& 1.6320$ \times 10^{-7} $& Khare \& Kumar (1977)\\
& $ \rightarrow$ N+ N &133.5 & 4.4683$ \times 10^{-6} $ &3.5474$ \times 10^{-6} $ &2.8628$ \times 10^{-6} $ &6.7673$ \times 10^{-7} $& 3.2275$ \times 10^{-7} $&"  \\
& $ \rightarrow$ N$_2^+$ + e$^-$ &36.3 & 1.6433$ \times 10^{-5}$ &1.3046$ \times 10^{-5} $&1.0529$ \times 10^{-5} $&2.4886$ \times 10^{-6} $&1.1870$ \times 10^{-6} $& "  \\
\tableline
He& $ \rightarrow$ He$^+$ + e$^-$ &46.3& 1.2884$ \times 10^{-5} $ &1.0229$ \times 10^{-5} $ &8.2549$ \times 10^{-6} $&1.9511$ \times 10^{-6} $&9.3054$ \times 10^{-7} $& "  \\
Ne& $ \rightarrow$ Ne$^+$ + e$^-$ &36.4& 1.6387$ \times 10^{-5} $ &1.3010$ \times 10^{-5} $ &1.0500$ \times 10^{-5} $&2.4819$ \times 10^{-6} $& 1.1837$ \times 10^{-6} $&"  \\
Ar& $ \rightarrow$ Ar$^+$ + e$^-$ &26.2& 2.2767$ \times 10^{-5} $ &1.8075$ \times 10^{-5} $ &1.4588$ \times 10^{-5} $&3.4481$ \times 10^{-6} $& 1.6445$ \times 10^{-6} $&"  \\
\tableline
\enddata

\tablenotetext{a}{The rate is expressed in a Arrhenius form k$_{C}$: {\rm A $\times \exp(-3386.5/T)$}. See Cherchneff \& Dwek (2010) for details.}
\end{deluxetable}
\clearpage

\clearpage




\clearpage

\clearpage


\clearpage
 

\clearpage

\clearpage


\clearpage


\clearpage
\clearpage


\clearpage
\clearpage
\clearpage


\begin{deluxetable}{lccccccccccccc}
\rotate
\tabletypesize{\scriptsize}
\tablecaption{Initial (post-explosive) elemental mass yields (in \Ms) as a function of progenitor mass and ejecta zone. The zone extention (in mass coordinates \Ms), the mean molecular weight (in g cm$^{-3}$), and the C/O ratio are also given for each zone.\tablenotemark{a}\label{tbl-1}}
 \tablewidth{0pt}

\tablehead{
\colhead{Zone } & \colhead{$\mu$(gas)} & \colhead{C/O} & \colhead{He} &\colhead{C} & \colhead{O}& \colhead{Ne}& \colhead{Mg}& \colhead{Al}& \colhead{Si}& \colhead{S} &\colhead{Ar}& \colhead{Fe}&  \colhead{Ni}  \\
}
\startdata
\multicolumn{14}{c}{\bf{12 \Ms}} \\
\hline
1A ($1.7-1.76$ \Ms) & 32.02 & 0.215 & 0 & 9.76(-8) & 6.71(-7) & 3.05(-7) & 7.32(-6) & 1.16(-5) & 2.50(-2) & 2.38(-2) & 3.29(-3) & 4.76(-3)  & 3.54(-5) \\
1B ($1.76-1.89$ \Ms) & 17.38 & 3.68(-4) & 0 & 2.86(-5) & 0.104 & 1.30(-5) & 5.72(-3) & 4.16(-4) & 1.56(-2) & 2.60(-3) & 7.41(-5) & 5.07(-5) & 1.24(-4) \\
2 ($1.89-2.03$ \Ms) & 17.3 & 7.87(-3) & 0 & 5.46(-4) & 9.24(-2) & 3.50(-2) & 8.68(-3) & 5.60(-4) & 6.02(-4) & 4.20(-5) & 1.19(-5) & 1.29(-4)& 0 \\
3 ($2.03-2.19$ \Ms) & 15.28 & 0.34 & 0 & 3.04(-2) & 0.118 & 7.36(-3) & 2.72(-3) & 2.08(-4) & 1.47(-4) & 3.52(-5) & 1.42(-5) & 8.16(-5) &  0 \\
4 ($2.19-2.35$ \Ms) & 4.91 & 15.39 & 0.117 & 3.68(-2) & 3.20-(3) & 2.08(-3) & 8.00(-4) & 9.28(-6) & 1.47(-4) & 4.96(-5) & 1.31(-5) & 1.76(-4) & 0 \\
5 ($2.35-3.27$ \Ms) & 4.05 & 1.27 & 0.911 & 1.93(-4) & 2.02(-4) & 1.01(-2) & 6.62(-4) & 6.75(-5) & 7.54(-4) & 3.86(-4) & 1.01(-4) & 1.29(-3) & 0 \\
\hline
\multicolumn{14}{c}{\bf{15 \Ms}} \\
\hline
1A ($1.79-1.88$ \Ms) & 35.49& 5.9(-2) & 0 & 1.45(-7) & 3.30(-6) & 0 & 1.39(-5) & 1.98(-5) & 3.19(-2) & 1.96(-2) & 4.02(-3) & 1.65(-2) & 2.80(-4) \\
1B ($1.88-1.98$ \Ms) & 20.89& 2.1(-3) & 0 & 6.91(-6) & 4.36(-2) & 1.05(-5) & 3.92(-4) & 4.97(-5) & 3.12(-2) & 1.25(-2) & 7.38(-4) & 1.25(-4) & 1.24(-7) \\
2 ($1.98-2.27$ \Ms) & 17.17 & 5.50(-3) & 0 & 9.26(-4) & 0.225 & 1.51(-2) & 1.60(-2) & 2.10(-3) & 2.10(-2) & 2.52(-3) & 4.06(-5) & 2.30(-5) & 0 \\
3 ($2.27-2.62$ \Ms) & 17.12 & 1.60(-2) & 0 & 2.77(-3) & 0.234 & 7.76(-2) & 1.75(-2) & 1.92(-3) & 1.76(-3) & 6.84(-5) & 1.72(-5) & 3.38(-5)  & 0 \\
4A($2.62-2.81$ \Ms) &14.99 & 0.367 & 6.06(-6) & 4.04(-2) & 0.147 & 2.97(-3) & 1.64(-4) & 2.34(-4) & 7.08(-5) & 3.47(-5) & 9.56(-6) & 2.15(-5)  & 0 \\
4B ($2.81-3.04$ \Ms) & 10.66 & 0.735 & 3.08(-2) & 6.16(-2) & 0.112 & 1.40(-2) & 7.11(-4) & 1.91(-5) & 1.05(-4) & 4.39(-5) & 8.64(-6) & 4.01(-5)  & 0 \\
5 ($3.04-3.79$ \Ms) & 4.14 & 21.3 & 0.705 & 2.72(-2) & 1.66(-3) & 1.19(-3) & 3.86(-4) & 5.25(-5) & 4.84(-4) & 2.91(-4) & 1.20(-5) & 8.40(-4) &  0 \\
6 ($3.79-4.14$ \Ms) & 4.05 & 1.18 & 0.341 & 9.13(-5) & 9.58(-5) & 5.48(-4) & 1.79(-4) & 2.43(-5) & 2.27(-4) & 1.37(-4) & 5.34(-6) & 4.06(-4) &  0 \\
\hline
\multicolumn{14}{c}{\bf{19 \Ms}} \\
\hline
1A ($1.77-1.88$ \Ms) & 35.35 & 0.156 & 1.39(-6) & 8.10(-8) & 6.89(-7) & 0 & 1.69(-5) & 2.46(-5) & 3.77(-2) & 2.26(-2) & 4.52(-3) & 2.47(-2) &  3.25(-4) \\
1B ($1.88-2.18$ \Ms) & 22.47 & 1.30(-3) & 0 & 1.15(-4) & 0.118 & 1.05(-4) & 8.79(-4) & 2.18(-4) & 9.88(-2) & 5.59(-2) & 1.54(-2) & 3.10(-3)  & 6.19(-6) \\
2 ($2.18-3.86$ \Ms) & 16.89 & 6.54(-2) & 0 & 5.92(-2) & 1.16 & 0.34 & 8.41(-2) & 9.12(-3) & 1.51(-2) & 1.28(-3) & 1.21(-4) & 7.54(-4) &  0 \\
3 ($3.86-4.00$ \Ms) & 15.11 & 0.36 & 0 & 2.89(-2) & 0.107 & 2.81(-3) & 2.01(-3) & 1.63(-5) & 1.11(-4) & 3.19(-5) & 8.94(-6) & 7.82(-5)  & 0 \\ 
4 ($4.00-4.49$ \Ms) & 10.32 & 0.64 & 7.68(-2) & 0.126 & 0.263 & 1.32(-2) & 5.44(-3) & 5.30(-5) & 3.75(-4) & 1.40(-4) & 3.47(-5) & 3.76(-4)  & 0 \\
5 ($4.49-5.26$ \Ms) & 4.12 & 3.93 & 0.743 & 1.27(-2) & 4.30(-3) & 1.07(-2) & 5.01(-4) & 5.99(-5) & 5.53(-4) & 3.21(-4) & 7.04(-5) & 9.83(-4)  & 0 \\
6 ($5.26-5.62$ \Ms) & 4.06 & 1.8 & 0.352 & 1.26(-4) & 9.24(-5) & 5.69(-4) & 2.27(-4) & 3.19(-5) & 2.55(-4) & 1.50(-4) & 3.32(-5) & 4.55(-4)  & 0 \\
\hline
\multicolumn{14}{c}{\bf{25 \Ms}} \\
\hline
1A ($2.1-2.33$ \Ms) & 34.26 & -- & 2.91(-6) & 0 & 0 & 0 & 4.91(-5) & 4.80(-5) & 8.92(-2) & 5.15(-2) & 9.93(-3) & 3.20(-2)  & 5.43(-4) \\
1B ($2.33-2.51$ \Ms) & 26.18 & 4.03(-4) & 0 & 8.18(-6) & 2.84(-2) & 1.71(-5) & 5.90(-5) & 6.46(-5) & 8.87(-2) & 4.34(-2) & 8.53(-3) & 3.81(-4) & 1.36(-6) \\
2 ($2.51-2.98$ \Ms) & 19.34 & 4.18(-4) & 0 & 8.70(-5) & 0.278 & 6.07(-5) & 7.22(-3) & 6.95(-4) & 0.116 & 3.74(-2) & 9.17(-3) & 1.32(-4) & 0 \\
3 ($2.98-5.69$ \Ms) & 17.01 & 2.32(-2) & 0 & 4.3(-2) & 1.95 & 0.44 & 0.129 & 2.12(-2) & 5.81(-2) & 8.29(-3) & 1.09(-3) & 1.20(-4)  & 0 \\
4A ($5.69-6.22$ \Ms) & 15.04 & 0.33 & 0 & 0.102 & 0.406 & 5.03(-3) & 3.02(-4) & 7.15(-5) & 1.76(-4) & 8.94(-5) & 3.16(-5) & 3.95(-5)  & 0 \\
4B ($6.22-7.11$ \Ms) & 12.40 & 0.49 & 6.17(-2) & 0.231 & 0.525 & 4.64(-2) & 2.61(-3) & 1.36(-4) & 3.18(-4) & 1.40(-4) & 4.98(-5) & 7.56(-5) & 0 \\
5 ($7.11-8.07$ \Ms) & 4.05 & 35.9 & 0.919 & 1.30(-2) & 4.82(-4) & 1.50(-3) & 4.78(-4) & 7.21(-5) & 6.08(-4) & 3.62(-4) & 8.61(-5) & 1.00(-3)  & 0 \\
6 ($8.07-8.30$ \Ms) & 4.05 & 1.57 & 0.232 & 6.89(-5) & 5.83(-5) & 3.71(-4) & 1.22(-4) & 1.77(-5) & 1.54(-4) & 9.35(-5) & 2.13(-5) & 2.76(-4)  & 0 \\

\enddata
\tablenotetext{a}{Data for the 12 \Ms~progenitor are from \citet{woos07} while data for the 15 \Ms, 19 \Ms, and 25 \Ms~are from \citet{rau02}.}
\end{deluxetable}
\clearpage

\begin{deluxetable}{c|cc|cc|cc|cc|cc|cc|cc}
\rotate
\tabletypesize{\scriptsize}
\tablecaption{Gas temperature and number density variation with post-explosion time for each zone of the SN ejecta with 15 \Ms\ progenitor\tablenotemark{a}.}

 \tablewidth{0pt}
\tablehead{

15 \Ms & \multicolumn{2}{c}{Zone 1A} & \multicolumn{2}{c}{Zone 1B} & \multicolumn{2}{c}{Zone 2} & \multicolumn{2}{c}{Zone 3} & \multicolumn{2}{c}{Zone 4A} & \multicolumn{2}{c}{Zone 4B} &  \multicolumn{2}{c}{Zone 5}\label{tbl-2}
}
\startdata
 Time (days) & T & n$_{gas}$ & T & n$_{gas}$ & T & n$_{gas}$ & T & n$_{gas}$& T & n$_{gas}$ & T & n$_{gas}$ & T & n$_{gas}$ \\
\hline\\

100 & 12000 & 1.8(11) & 11600 & 3.1(11) & 10400 & 3.7(11) & 8779 & 3.8(11) & 7980 & 4.3(11) & 7580  & 6.1(11) & 6490 & 1.6(12)  \\
300 & 3006 & 6.7(9) & 2906 & 1.1(10) & 2605 & 1.4(10)  & 2199 & 1.4(10) & 1998 & 1.6(10) & 1899  & 2.3(10) & 1626 & 5.9(10) \\
600 & 1255 & 8.3(8) & 1213 & 1.4(9) & 1088 & 1.7(9) & 918 & 1.8(9) & 835 & 2.0(9) & 793 & 2.8(9) & 679 & 7.4(9) \\
900 & 753 & 2.5(8) & 728 & 4.3(8) & 653 & 5.1(8)&  551 & 5.2(8) & 501 & 5.9(8) & 476  & 8.4(8) & 407 & 2.2(9) \\
1200 & 524 & 1.0(8) & 507 & 1.8(8) & 454 & 2.1(8)&  383 & 2.2(8) & 349 & 2.5(8) & 331 & 3.5(8) & 283 & 9.3(8) \\
1500 & 396 & 5.3(7) & 382 & 9.2(7) & 343 & 1.1(8)&  289 & 1.1(8) & 263 & 1.3(8) & 250 & 1.8(8) & 214 & 4.7(8) \\

\tablenotetext{a}{The temperature T is in Kelvin and the gas number density n$_{gas}$ in cm$^{-3}$.}
\enddata

\end{deluxetable}

\begin{deluxetable}{l l l l l l l l l l }
\tabletypesize{\scriptsize}
\tablecaption{Chemical species and dust clusters included in the chemical model of the SN ejecta.\label{tbl-3}}             
\label{tab3}    
 \tablewidth{0pt}
\tablehead{
\multicolumn{9}{c}{\bf{Elements}} \\
}
\startdata
O & Si & S & C & Mg & Al& Fe & He & Ne & Ar \\ 
\hline
\multicolumn{9}{c}{\bf{Ions}} \\
\hline
O$^+$ & Si$^+$ & S$^+$ & C$^+$ & Mg$^+$ & Al$^+$& Fe$^+$ & He$^+$ & Ne$^+$ & Ar$^+$\\ 
SiO$^+$ & CO$^+$ & O$_2^+$ & SO$^+$ & & & & & \\
\hline
\multicolumn{9}{c}{\bf{Molecules}} \\
\hline
O$_2$ & CO & SiO & SO & NO& AlO & FeO & MgO & CO$_2$ & \\ 
CN & CS &SiS & SiC &  FeS & MgS & S$_2$ & N$_2$ & \\
\hline
\multicolumn{9}{c}{\bf{Clusters}} \\
\hline
C$_2$ & C$_3$ & C$_4$& C$_5$ & C$_6$ & C$_7$ & C$_8$ & C$_9$ & C$_{10}$ & \\
Si$_2$ & Si$_3$ & Si$_4$& Mg$_2$ & Mg$_3$ & Mg$_4$& Fe$_2$ & Fe$_3$ & Fe$_4$ & \\
Si$_2$C$_2$ & Mg$_2$S$_2$ & Mg$_3$S$_3$ & Mg$_4$S$_4$& Fe$_2$S$_2$ & Fe$_3$S$_3$ & Fe$_4$S$_4$& \\
Si$_2$O$_2$& Si$_3$O$_3$& Si$_4$O$_4$&Si$_5$O$_5$ & SiO$_2$ & Si$_2$O$_3$&  Si$_3$O$_4$ & Si$_4$O$_5$ &  \\ 
MgSi$_2$O$_3$ & MgSi$_2$O$_4$ & Mg$_2$Si$_2$O$_4$ & Mg$_2$Si$_2$O$_5$ & Mg$_2$Si$_2$O$_6$ & Mg$_3$Si$_2$O$_6$ & Mg$_3$Si$_2$O$_7$ & Mg$_4$Si$_2$O$_7$ & Mg$_4$Si$_2$O$_8$ \\
Mg$_2$O$_2$ & Mg$_3$O$_3$ & Mg$_4$O$_4$ &Fe$_2$O$_2$ & Fe$_3$O$_3$ & Fe$_4$O$_4$ &&& \\
\enddata
\end{deluxetable}

 \begin{deluxetable}{cccccccccc}
\tabletypesize{\scriptsize}
\tablecaption{Masses of molecules and upper limit of dust masses (both in \Ms) at 1500 days for the 15 \Ms~model, and two values of the \Ni~mass (0.075 \Ms~and 0.01 \Ms). Efficiencies are the molecule- or dust-to-gas mass ratio in each zone and for the total ejected zones of the He core.\label{tbl-4}}

\tablewidth{0pt}
\tablehead{

\colhead{Ejected zones} & \colhead{Zone 1A} & \colhead{Zone 1B} & \colhead{Zone 2} & \colhead{Zone 3} & \colhead{Zone 4A} &\colhead{Zone 4B} &  \colhead{Zone 5} &  \colhead{Zone 6} &\colhead{{\bf Total}}}

\startdata
Zone Mass(\Ms) & 9.6(-2) & 9.5(-2) & 0.292 & 0.347 & 0.195 & 0.225 & 0.75 & 0.347 & \bf{2.35} \\
Major Elements & Si/S/Fe & Si/O & O/Mg/Si & O/Ne/Mg & O/C & He/O/C & He/C & He/N & \\ 

\hline
\hline
\multicolumn{10}{c}{\bf{MOLECULES ($M(^{56}Ni$) = 0.075 \Ms})} \\
\hline
SiO & 4.4(-7) & 3.0(-7) & 4.4(-7) & 1.3(-7) & 3.1(-8) & 2.0(-8) & ... & ... & 1.4(-6) \\
$O_{2}$ & ... & 2.8(-5) & 0.15 & 0.16 & 6.2(-2) & 4.9(-3) & ... & ... & 0.38 \\
CO & 7.5(-7) & 1.7(-5) & 2.2(-3) & 6.6(-3) & 9.5(-2) & 0.14 & 2.9(-3) & ... & 0.25 \\
SO & ... & 1.5(-2) & 3.8(-3) & 1.0(-4) & 1.1(-4) & ... & ... & ... & 1.9(-2) \\
SiS & 4.3(-2) & 2.1(-7) & ... & ... & ... & ... & ... & ... & 4.3(-2) \\
\hline
Total Mass(\Ms) & 4.3(-2) & 1.5(-2) & 0.156 & 0.167 & 0.157 & 0.145 & 2.9(-3) & 0 & \bf{0.69} \\
Efficiency & 44.8\% & 15.8\% & 53.4\% & 48.1\% & 80.5\% & 64.7\% & 0.4\% & 0\% & \bf{29.4\%} \\
\hline
\multicolumn{10}{c}{\bf{DUST ($M(^{56}Ni$) = 0.075 \Ms})}\\
\hline
Forsterite & ... & 5.3(-4) & 4.4(-3) & 5.9(-4) & 2.7(-5) & 2.5(-5) & ... & ... & 5.6(-3) \\
Silica & ... & 6.0(-5) & 5.1(-5) & ... & ... & ... & ... & ... & 1.1(-4) \\
Alumina & ... & 1.2(-5) & 4.0(-3) & 3.7(-3) & 4.5(-5) & 3.5(-5) & ... & ... & 7.8(-3) \\
Pure Iron & 1.2(-4) & ... & ... & ... & ... & ... & ... & ... & 1.2(-4) \\
Iron Sulfide & 2.1(-6) & ... & ... & ... & ... & ... & ... & ... & 2.1(-6) \\
Pure Silicon & 3.9(-4) & ... & ... & ... & ... & ... & ... & ... & 3.9(-4) \\
Pure Magnesium & ... & ... & 2.2(-4) & 2.5(-4) & ... & ... & ... & ... & 4.7(-4) \\
Carbon & ... & ... & ... & ... & ... & ... & 2.3(-2) & ... & 2.3(-2) \\
Silicon Carbide & ... & ... & ... & ... & ... & ... & 6.1(-4) & ... & 6.1(-4) \\
\hline
Total Mass(\Ms) & 5.1(-4) & 6.0(-4) & 8.7(-3) & 4.5(-3) & 7.2(-5) & 6.0(-5) & 2.4(-2) & ... & \bf{0.038}\\
Efficiency(\%) & 0.53\% & 0.63\% & 3.0\% & 1.3\% & 3.7(-2)\% & 2.7(-2)\% & 3.2\% & ... & \bf{1.62\%}\\
\hline
\hline
\multicolumn{10}{c}{\bf{MOLECULES ($M(^{56}Ni$) = 0.01 \Ms}) }\\
\hline
SiO & 3.8(-7) & 2.1(-7) & 4.3(-7) & 1.3(-7) & 3.1(-8) & 2.0(-8) & ... & ... & 1.2(-6) \\
$O_{2}$ & ... & 3.4(-4) & 0.16 & 0.18 & 6.8(-2) & 8.3(-3) & ... & ... & 0.42 \\
CO & 7.6(-7) & 1.7(-5) & 2.2(-3) & 6.6(-3) & 9.5(-2) & 0.14 & 2.8(-3) & ... & 0.25 \\
SO & ... & 1.9(-2) & 3.9(-3) & 1.0(-4) & 8.0(-5) & ... & ... & ... & 2.3(-2) \\
SiS & 4.3(-2) & 4.0(-7) & ... & ... & ... & ... & ... & ... & 4.3(-2) \\
\hline
Total Mass(\Ms) & 4.3(-2) & 1.9(-2) & 0.166 & 0.187 & 0.163 & 0.15 & 2.8(-3) & 0 & \bf{0.73} \\
Efficiency & 44.8\% & 20.0\% & 56.8\% & 53.9\% & 83.6\% & 66.7\% & 0.4\% & 0\% & \bf{31.1\%} \\
\hline
\multicolumn{10}{c}{\bf{DUST ($M(^{56}Ni$) = 0.01 \Ms}) }\\
\hline
Forsterite & ... & 6.7(-4) & 2.3(-2) & 1.8(-3) & 1.2(-4) & 7.2(-5) & ... & ... & 2.6(-2)\\
Silica & ... & 6.0(-5) & 4.1(-5) & ... & ... & ... & ... & ... & 1.1(-4) \\
Alumina & ... & 6.4(-5) & 4.0(-3) & 3.7(-3) & 4.5(-5) & 3.5(-5) & ... & ... & 7.9(-3) \\
Pure Iron & 1.2(-4) & ... & ... & ... & ... & ... & ... & ... & 1.2(-4) \\
Iron Sulfide & 3.1(-6) & ... & ... & ... & ... & ... & ... & ... & 3.1(-6) \\
Pure Silicon & 3.8(-4) & ... & ... & ... & ... & ... & ... & ... & 3.8(-4) \\
Pure Magnesium & ... & ... & 6.8(-5) & 3.4(-4) & ... & ... & ... & ... & 4.1(-4) \\
Carbon & ... & ... & ... & ... & ... & ... & 2.4(-2) & ... & 2.4(-2) \\
Silicon Carbide & ... & ... & ... & ... & ... & ... & 5.0(-4) & ... & 5.0(-4) \\
\hline
Total Mass(\Ms) & 5.0(-4) & 7.9(-4) & 2.7(-2) & 5.8(-3) & 1.7(-4) & 1.1(-4) & 2.4(-2) & ... & \bf{0.059}\\
Efficiency& 0.52\% & 0.83\% & 9.2\% & 1.7\% & 8.7(-2)\% & 4.9(-2)\% & 3.2\% & ... & \bf{2.5\%}\\

\enddata

\end{deluxetable}
\clearpage


\begin{deluxetable}{lcccl}
\tabletypesize{\scriptsize}
\tablecaption{\Ni~mass (in \Ms), and low and high limits on progenitor mass (in \Ms) for a sample of Type II-P SNe. The progenitor masses of well-studied SN remnants with Type II SN progenitors are also indicated.\label{tbl-5}}             
\tablewidth{0pt}
\tablehead{
\colhead{Name} & \colhead{\Ni~mass} & \colhead{Low Limit } & \colhead{High Limit } & \colhead{Reference}}
\startdata
SN1999em& 0.02 & 12 & 14 & \citet{elma03}\\
SN2003gd &  0.016& 8 & 12 &  \citet{smart04,hen05} \\
SN2004dj & 0.095 & 12 & 20 & \citet{wan05}; \citet{vin09} \\
SN2004et & 0.068 & 23 & 25 & \citet{ko09} \\
SN2005af & 0.027 & 13 &15 & \citet{ko06} \\
SN2005cs & 0.003 & 10 & 15& \citet{pas09} \\
SN2007od & 0.02& 10& 11& \citet{an11}; \citet{ins11}\\
SN2009bw & 0.022 & 11 & 15 & \citet{ins12} \\
SN2009js & 0.007 & 6 & 16 & \cite{gan13} \\
SN2011ht\tablenotemark{a} & 0.01 & 8 & 10 & \citet{mau12}  \\
\hline
\multicolumn{5}{c}{SN remnants}
\\
\hline
SN1987A & 0.075 & 14 & 20 &  \citet{woos88}, \cite{smart09}\\ 
Cas A &--& 18& 20&\citet{krau08} \\
The Crab & --&8 & 12& \citet{dav85,alp08} \\

\enddata
\tablenotetext{a}{A high-mass progenitor ($\ge$ 25 \Ms) with substantial fallback of the ejecta is also possible for this object}
\end{deluxetable}

\begin{deluxetable}{ccccccccc}
\tabletypesize{\scriptsize}
\tablecaption{Masses of molecules and upper limit of dust masses (both in \Ms) at 1500 days for the 19 \Ms~progenitor and a \Ni~mass of 0.075 \Ms. Efficiencies are the molecule- or dust-to-gas mass ratio in each zone and for the total ejected zones of the He core.\label{tbl-6} }

 \tablewidth{0pt}
\tablehead{

\colhead{Ejected zones} & \colhead{Zone 1A} & \colhead{Zone 1B} & \colhead{Zone 2} & \colhead{Zone 3} & \colhead{Zone 4}  &  \colhead{Zone 5} &  \colhead{Zone 6} &\colhead{{\bf Total}}}
\startdata

Zone Mass(\Ms) & 0.11 & 0.302 & 1.68 & 0.141 & 0.486 & 0.774 & 0.358 & \bf{3.85} \\
Major Elements & Si/S/Fe & Si/O & O/Ne/Mg & O/C & He/O/C & He/C & He/N & \\ 

\hline
\multicolumn{9}{c}{\bf{MOLECULES}} \\
\hline
SiO & 3.9(-7) & 2.0(-7) & 1.9(-6) & 7.9(-9) & 7.2(-9) & ... & ... & 2.5(-6) \\
O$_{2}$ & ... & 1.0(-5) &  0.69 & 4.7(-2) & 3.4(-2) & ... & ... & 0.77 \\
CO & 1.5(-8) & 2.7(-4) &  0.13 & 6.7(-2) & 0.29 & 7.2(-3) & ... & 0.50 \\
SO & ... & 2.3(-2) & 1.9(-3) & 1.4(-4) & 1.0(-7) & ... & ... & 2.5(-2) \\
SiS & 4.4(-2) & 8.6(-5) & ... & ... & ... & ... & ... & 4.4(-2) \\
\hline
Total Mass(\Ms) & 4.4(-2) & 2.3(-2) & 0.82 & 0.115 & 0.29 & 7.2(-3) & 0 & \bf{1.33} \\
Efficiency & 40.0\% & 7.6\% & 48.8\% & 81.6\% & 59.7\% & 0.93\% & 0\% & \bf{33.9\%} \\
\hline
\multicolumn{9}{c}{\bf{DUST}} \\
\hline
Forsterite & ... & 1.7(-3)  & 6.5(-3) & 1.6(-4) & 2.5(-4) & ... & ... & 8.6(-3) \\
Silica & ... & 3.0(-4) & 2.3(-4) & ... & ... & ... & ... & 5.3(-4) \\
Alumina & ... & 8.5(-6) & 1.79(-2) & 3.1(-5) & 1.0(-4) & ... & ... & 1.8(-2) \\
Pure Iron & 2.0(-4) & ... & ... & ... & ... & ... & ... & 2.0(-4) \\
Iron Sulfide & 5.4(-8) & ... & ... & ... & ... & ... & ... & 5.4(-8) \\
Pure Silicon & 3.6(-4) & ...& ... & ... & ... & ... & ... & 3.6(-4) \\
Pure Magnesium & ... & ... & 9.9(-4) & ... & ... & ... & ... & 9.9(-4) \\
Carbon & ... & ... & ... & ... & ... & 5.5(-3) & ... & 5.5(-3) \\
Silicon Carbide & ... & ... & ... & ... & ... & 2.6(-4) & ... & 2.6(-4) \\
\hline
Total Mass(\Ms) & 5.6(-4) & 2.0(-3) & 2.6(-2) & 1.9(-4) & 3.5(-4) & 5.8(-3) & 0 & \bf{0.035}\\
Efficiency & 0.51\% & 0.66\% & 1.6\% & 0.13\% & 7.2(-2)\% & 0.71\% & 0\% & \bf{0.91\%}\\

\enddata

\end{deluxetable}

\begin{deluxetable}{lccccccccc}
\tabletypesize{\scriptsize}
\tablenum{7}
\tablecaption{Masses of molecules and upper limit of dust masses (both in \Ms) at 2000 days for the 25 \Ms~progenitor with a \Ni~mass of 0.075 \Ms. Efficiencies are the molecule- or dust-to-gas mass ratio in each zone and for the total ejected zones of the He core.\label{tbl-7} }                                                                                             
\tablewidth{0pt}
\tablehead{

\colhead{Ejected zones} & \colhead{Zone 1A} & \colhead{Zone 1B} & \colhead{Zone 2} & \colhead{Zone 3} & \colhead{Zone 4A} & \colhead{Zone 4B} & \colhead{Zone 5} & \colhead{Zone 6} & \colhead{{\bf Total}}}
\startdata
Zone Mass(\Ms) & 0.233 & 0.181 & 0.463 & 2.72 & 0.526 & 0.89 & 0.956 & 0.236 & \bf{6.21} \\
Major Elements & Si/S/Fe & Si/O/S & O/Mg/Si & O/Ne/Mg & O/C & He/O/C & He/C & He/N & \\ 
\hline
\multicolumn{10}{c}{\bf MOLECULES} \\
\hline
SiO & 5.5(-8) & 2.3(-7) & 1.3(-7) & 1.2(-6) & 2.5(-7) & 3.9(-7) & ... & ... & 2.2(-6) \\
$O_{2}$ & ... & ... & 0.10 & 1.4 & 0.18 & 7.5(-2) & ... & ... & 1.76 \\
CO & ... & 2.0(-5) & 2.1(-4) & 7.9(-2) & 0.24 & 0.54 & 7.6(-4) & ... & 0.86 \\
SO & ... & 1.9(-7) & 5.6(-2) & 1.2(-2) & 2.7(-4) & ... & ... & ... & 6.8(-2) \\
SiS & 0.12 & 1.6(-2) & 1.5(-6) & ... & ... & ... & ... & ... & 0.1 \\
\hline
Total Mass (\Ms) & 0.12 & 1.6(-2) & 0.156 & 1.49 & 0.42 & 0.62 & 7.6(-4) & 0 & \bf{2.82} \\
Efficiency(\%) & 51.5 \% & 8.9 \% & 33.6 \% & 54.8 \% & 79.8 \% & 70.7 \% & 7.9(-2) \% &  & \bf{45.4\%} \\
\hline
\multicolumn{10}{c}{\bf DUST} \\
\hline
Forsterite - Mg$_2$SiO$_4$ & ... & 1.6(-5) & 1.9(-2) & 1.2(-2) & 1.1(-4) & 1.8(-4) & ... & ... & 3.2(-2) \\
Silica - SiO$_2$ & ... & ... & 3.1(-4) & 6.0(-4) & ... & ... & ... & ... & 9.1(-4) \\
Alumina - Al$_2$O$_3$ & ... & 1.3(-3) & 3.9(-2) & 1.3(-4) & 2.6(-4) & 3.5(-5) & ... & ... & 4.1(-2) \\
Pure Iron - Fe & 5.7(-4) & ... & ... & ... & ... & ... & ... & ... & 5.7(-4) \\
Iron Sulfide - FeS & 6.6(-7) & ... & ... & ... & ... & ... & ... & ... & 6.6(-7) \\
Pure Silicon - Si & 2.5(-3) & ... & ... & ... & ... & ... & ... & ... & 2.5(-3) \\
Pure Magnesium - Mg  & ... & ... & ... & 2.2(-3) & ... & ... & ... & ... & 2.2(-3) \\
Carbon - C$_{10}$& ... & ... & ... & ... & ... & ... & 1.0(-2) & ... & 1.0(-2) \\
Silicon Carbide - SiC & ... & ... & ... & ... & ... & ... & 8.2(-4) & ... & 8.2(-4) \\
\hline
Total Mass (\Ms) & 3.1(-3) & 1.4(-3) & 5.8(-2) & 1.5(-2) & 3.7(-4) & 2.2(-4) & 1.1(-3) & 0 & \bf{0.09}\\
Efficiency & 1.33 \% & 0.77 \% & 12.5 \% & 0.55 \% & 7.0(-2) \% & 2.5(-2) \% & 0.11 \% &  & \bf{1.3 \%} \\
\enddata

\end{deluxetable}

\begin{deluxetable}{lccccccc}
\tabletypesize{\scriptsize}
\tablenum{8}
\tablecaption{Masses of molecules and upper limit of dust masses (both in \Ms) at 1500 days for the 12 \Ms~progenitor with a \Ni~mass of 0.001 \Ms. Efficiencies are the molecule- or dust-to-gas mass ratio in each zone and for the total ejected zones of the He core.\label{tbl-8} }

 \tablewidth{0pt}
\tablehead{

\colhead{Ejected zones} & \colhead{Zone 1A} & \colhead{Zone 1B} & \colhead{Zone 2} & \colhead{Zone 3} & \colhead{Zone 4} & \colhead{Zone 5} & \colhead{{\bf Total}}}
\startdata
Zone Mass(\Ms) & 6.1(-2) & 0.13 & 0.14 & 0.16  & 0.16 & 0.92 & \bf{1.57} \\
Major Elements & Si/S/Fe & Si/O & O/Ne/Mg & O/C & He/C & He/N & \\ 
\hline
\multicolumn{8}{c}{\bf MOLECULES} \\
\hline
SiO & 3.0(-8) & 5.7(-8) & 6.0(-8)  & 7.0(-8) & ... & ... & 2.2(-7) \\
O$_{2}$ & ... & 6.8(-2) &  6.4(-2)  & 4.9(-2) & ... & ... & 0.18 \\
CO & 2.4(-7) & 6.5(-5) &  1.2(-3)  & 7.0(-2) & 5.5(-3) & ... & 7.7(-2) \\
SO & ... & 3.9(-3) & 6.2(-5) & 5.5(-9) ) & ... & ... & 4.0(-3) \\
SiS & 4.7(-2) & ... & ... & ... &  ... & ... & 4.7(-2) \\
\hline
Total Mass (\Ms) & 4.7(-2) & 7.2(-2) & 6.5(-2)  & 0.12 & 5.5(-3) & 0 & \bf{0.31} \\
Efficiency & 77 \% & 55.4 \%& 46.4 \%& 75 \% & 3.4 \% &  & \bf{19.7 \%} \\
\hline
\multicolumn{8}{c}{\bf DUST} \\
\hline
Forsterite - Mg$_2$SiO$_4$ & ... & 1.2(-2)  & 2.9(-4) & 2.2(-4) & ... &  ... & 1.3(-2) \\
Enstatite - MgSiO$_3$ &  & & & & & & \\
Silica - SiO$_2$ & ... & 5.2(-6) & 5.0(-6) & ... & ... &  ... & 1.2(-5) \\
Alumina - Al$_2$O$_3$ & ... & 7.9(-4) & 1.03(-3)  & 3.9(-4) & ... & ... & 2.2(-3) \\
Pure Iron - Fe& 2.3(-5) & ... & ... & ... & ... &  ... & 2.3(-5) \\
Iron Sulfide - FeS & 1.0(-7) & ... & ... & ... &  ... & ... & 1.0(-7) \\
Pure Silicon - Si& 8.3(-5) & ...& ... & ... &  ... & ... & 8.3(-5) \\
Pure Magnesium - Mg & ... & 8.1(-6) & 1.6(-4) & ... &  ... & ... & 1.7(-4) \\
Carbon - C$_{10}$& ... & ... & ... &  ... & 3.3(-2) & ... & 3.3(-2) \\
Silicon Carbide - SiC & ... & ... & ... & ... & 5.0(-6) & ... & 5.0(-6) \\
\hline
Total Mass (\Ms) & 1.1(-4) & 1.3(-2) & 1.5(-3) & 6.1(-4)  & 3.3(-2) & 0 & \bf{0.048}\\
Efficiency & 0.2\% & 10 \% & 1.1 \%& 0.4 \%  & 20.6 \%&  & \bf{3.1 \%} \\
\enddata

\end{deluxetable}
\clearpage


\begin{deluxetable}{lccccl}
\rotate
\tabletypesize{\scriptsize}
\tablenum{9}
\tablecaption{Condensation-time sequence and dust mass derived by existing dust formation models for Z = Z$_{solar}$ and Z = 0 metallicity.\tablenotemark{a} \label{tbl-9}}             
\tablewidth{0pt}
\tablehead{
\colhead{Model} & \colhead{Z}&\colhead{Fully Mixed} & \colhead{Progenitor }&\colhead{Dust Condensation-Time Sequence } & \colhead{Total Dust }\\
\colhead{ } & \colhead{}&\colhead{or Unmixed} & \colhead{ Mass} & \colhead{ } & \colhead{ Mass } }
\startdata
Kozasa et al. & Solar & FM & 19 \Ms& \al--\sif--\magn & -- \\
(1989)    & Solar  & U & 19 \Ms& Graphite& -- \\ 
    \hline
Kozasa et al.& Solar & U & 15 \Ms& AC--\al--\sif\ \& \sie--\sili--\magn--MgO--Si--FeS--Fe & 0.33 \Ms \\
(2009)    & Solar  & U & 20 \Ms&  --&  0.68 \Ms \\ 

\hline
Todini \& Ferrara& Solar & FM & 12 \Ms&  --    &0.20 \Ms\\
(2001) & Solar  & FM & 15 \Ms&  --    &0.45 \Ms\\
& Solar & FM &  20 \Ms&AC--\al--\magn--\sif--\sie  &0.70 \Ms\\ 
& Solar  & FM &  25 \Ms&  --    &1.00 \Ms\\ 
& 0 & FM &  15 \Ms&AC--\al--\magn--\sif--\sie  & 0.45 \Ms\\ 
& 0 & FM &  20 \Ms&AC  &0.08 \Ms\\ 
& 0 & FM &  25 \Ms&AC  &0.08 \Ms\\ 
\hline
Nozawa et al.  &  0& FM & 20 \Ms &\al--\sif--\sili--\magn& 0.73 \Ms \\
 (2003)&  0& U & 20 \Ms &AC--\al--\sif--MgO--\sili--Si--FeS--Fe& 0.57 \Ms \\
 \hline
Bianchi \& Schneider &  Solar& FM & 12 \Ms &-- & 0.12 \Ms \\
 (2007) &  Solar & FM & 15 \Ms &-- & 0.28 \Ms \\
 & Solar& FM  & 20 \Ms &AC--\al--\sif--\sili--\magn--\sie  & 0.40 \Ms \\
 &  Solar& FM  & 25 \Ms & --& 0.62 \Ms \\
\hline
\hline
Cherchneff \& Dwek \tablenotemark{b,c} & 0& FM & 20 \Ms & Mg--Si/Fe--\sili--\al& 0.16 \Ms \\
(2010) & 0 & U & 20 \Ms & \al--\sili--MgO--FeS--Si--Mg/Fe--AC & 0.10 \Ms \\
\hline
Sarangi \& Cherchneff \tablenotemark{b} & Solar & U & 12 \Ms & \al--\sif--\sili--Si/Mg/Fe--AC/SiC & 0.048 \Ms \\
 (this paper) & Solar & U & 15 \Ms & \sif/\sili--FeS--\al--Si--Fe--AC/SiC & 0.038 \Ms \\
  & Solar & U & 19 \Ms & \sif--\sili--Mg--Si--Fe--\al--AC/SiC& 0.035 \Ms \\
  & Solar & U & 25 \Ms & \sif--\sili--Si--\al--FeS--Fe/Mg--AC/SiC& 0.09 \Ms \\
\enddata
\tablenotetext{a}{Dust condensation-time sequences are only available from the literature for a set of progenitor masses.}
\tablenotetext{b}{The dust mass is derived from assuming 100 \% condensation of the dust clusters in grains. Mass values are then upper limits to the total dust masses that form in the ejecta.  }
\tablenotetext{c}{The formation of forsterite \& entastite is not modelled. \sili\ represents the generic class of silicates.} 
\end{deluxetable}

\end{document}